\documentclass[%11pt
preprint
]{elsarticle}

\journal{Journal of Computational and Applied Mathematics}

\usepackage{amsmath}
\usepackage{amssymb}
\usepackage{graphicx,color}
\usepackage{hyperref}

\newcommand{\newsection}[1]
{\section{#1}\setcounter{theorem}{0} \setcounter{equation}{0} \par\noindent}

\newtheorem{theorem}{Theorem}

\newcommand{\beq}{ \begin{equation} }
\newcommand{\eeq}{ \end{equation} }

\newcommand{\br}{{\mathbb R}}

\newcommand{\brn}{ { \mathbb{R}^n } }
\begin{document}

\begin{frontmatter}

  %%%%%  Title
  \title{
    On the numerical \textcolor{black}{experiments}
    of the Cauchy problem for semi-linear 
    Klein-Gordon equations in the de Sitter spacetime%\\
  }

  %%%%%  Authors
  \author[H]{Takuya Tsuchiya\corref{TT}}
  \author[Y]{Makoto Nakamura\corref{MN}}
  \cortext[TT]{%t-tsuchiya@aoni.waseda.jp
    \textcolor{black}{t-tsuchiya@hi-tech.ac.jp}
  }
  \cortext[MN]{nakamura@sci.kj.yamagata-u.ac.jp}
  \address[H]{
    \textcolor{black}{
      Center for Liberal Arts and Sciences,
      Hachinohe Institute of Technology,
      88-1, Obiraki, Myo, Hachinohe, Aomori 031-0814, JAPAN.
    %% School of Fundamendal Science and Engineering, 
    %% Waseda University,
      %% Shinjuku-ku Okubo 3-4-1, Tokyo 169-0072, JAPAN.
      }
  }
  \address[Y]{Faculty of Science, Yamagata University, 
    Kojirakawa-machi 1-4-12, Yamagata 990-8560, JAPAN.
  }

  \begin{abstract}
    The \textcolor{black}{computational analysis}
    of the Cauchy problem for semi-linear Klein-Gordon
    equations in the de Sitter spacetime is considered.
    Several simulations are performed to show the time-global behaviors 
    of the solutions of the equations in
    the spacetime based on the structure-preserving scheme.
    It is remarked that the sufficiently large Hubble constant yields the strong
    diffusion-effect which gives the long and stable simulations 
    for the defocusing semi-linear terms.
    The reliability of the simulations is confirmed by the preservation of the
    numerically modified Hamiltonian of the equations.
  \end{abstract}

\end{frontmatter}

\section{Introduction}
The mathematical structure of partial differential equations in non-flat
spacetimes is changed by the variations of curvatures since the differential
operators are influenced 
{by the metrics of the spacetimes.}
The de Sitter spacetime is one of the solutions of the Einstein gravitational
equation{s} with the cosmological constant in the vacuum. 
The space is expanding or contracting along the time in this spacetime.
To study the effects of the spatial variation, we consider the semilinear
field equation of the Klein-Gordon type, and we carry out the numerical
\textcolor{black}{experiments} of the solutions and their energy.
To investigate the partial differential equations with
\textcolor{black}{computational analysis},
we need to perform high precise and accurate simulations.
There are many methods to make discretized equations.
\textcolor{black}{
  %For performing numerical simulations of the Klein-Gordon equation in the de
  %Sitter spacetime, some results were repoted.
  %For instance, 
  The Adomian decomposition method \cite{Admian-1988-JMAA} can provide
  analytical approximation solutions of linear and nonlinear differential
  equations.
  Since this method dose not require the linearization, it would be
  appropriate to make dicretization of the nonlinear differential equations.
  It is reported that the simulations of the nonlinear Klein-Gordon equation
  in the flat spacetime with the method \cite{Basak-Ray-Bera-2009-CNS}.
  In \cite{Yazici-Sengul-2016-OP}, the differential transform
  method \cite{Zhou-1986} and the variational iteration
  method \cite{He-1999-IJNLM} are used to perform the nonlinear Klein-Gordon
  equation in the de Sitter spacetime.
  The former is one of the power series expansions.
  The virtue is easy to estimate the numerical errors because the truncation
  higher order terms are corresponding to the dominant numerical errors.
  Since the latter is classified as the method of Lagrange multipliers, it
  would be close to the exact solutions by repeating the iteration if the
  appropriate conditions are given.
  In addition, the results with the explicit fourth-order Runge-Kutta method
  are reported in \cite{Balogh-Banda-Yagdjian-2019-CNS}.
  The Runge-Kutta method is well-known discretized scheme for the nonlinear 
  differential equations.
  \\
  \indent The above methods are generic discretization schemes for the
  differential equations to calculate simulations.
  Therefore, the properties of the differential equations in continuous level
  often lost in the discretization, and it makes the numerical errors in
  the evolutions.
  On the other hand, we use the structure-preserving scheme (SPS) called as
  ``discrete variational derivative method'' in this paper.
  The concept of this method is to conserve the structures and the properties
  of the equations in continuous level.
  This method is widely used to derive the discrete equations of partial
  differential equations
  (see e.g., \cite{Furihata-1999-JComputPhys, Furihata-Matsuo-2010, Tsuchiya-Yoneda-2016-JSIAMLett}).
  To make discretized equations using SPS, we need the Lagrangian or
  the Hamiltonian\cite{Goldstein-Poole-Safko-CM, Wald}.
  This is because the derivations of the discretized equations with SPS are
  similar with the process of making the continuous equations using the
  variational principle.
  In comparison with the construction of the Lagrangian formulation, the one
  of the Hamiltonian formulation is clear.
  Especially, since the Hamiltonian means the total energy of the system, 
  this value is regarded as one of the constraints of the system.
  By monitoring the value in the evolution, we can judge whether the numerical simulations are success or not.
  Conservation of the constraints is one of the necessary conditions to perform
  successive simulations.
  If the constraints dose not conserve in the evolution, the simulations fail
(e.g. \cite{Tsuchiya-Yoneda-2016-Maxwell}).
Therefore, we use the Hamiltonian formulation in this paper.
}

We start from the introduction of the de Sitter spacetime.
Let $n\ge1$ be the spatial dimension, $c>0$ be the speed of light.
In the following, Greek letters $\alpha, \beta,\gamma,\dots$
run from $0$ to
$n$, Latin letters $j,k,\ell,\dots$ run from $1$ to $n$.
Let $g_{\alpha\beta}dx^\alpha dx^\beta$ be the metric in $\mathbb{R}^{1+n}$.
We denote by $(g_{\alpha\beta})_{0\le \alpha,\beta\le n}$ the
$(1+n)\times (1+n)$-matrix whose $(\alpha,\beta)$-component is given by
$g_{\alpha\beta}$.
Put $g :={\rm det}(g_{\alpha\beta})_{0\le \alpha,\beta\le n}$, and let
$(g^{\alpha\beta})_{0\le \alpha,\beta\le n}$ be the inverse matrix of
$(g_{\alpha\beta})_{0\le \alpha,\beta\le n}$.
We use the Einstein rule for the sum of indices of the tensors, for example,
${T^\alpha}_\alpha:=\sum_{\alpha=0}^n {T^\alpha}_\alpha$ and
${T^i}_i:=\sum_{i=1}^n {T^i}_i$.
The change of upper and lower indices of the tensors is done by
$g_{\alpha\beta}$ and $g^{\alpha\beta}$, for example,
${T^\alpha}_\beta:=g^{\alpha\gamma}T_{\gamma\beta}$.

For a stress-energy tensor ${T^\alpha}_\beta$, the $(1+n)$-dimensional
Einstein equation is defined by
\begin{equation}
  \label{Einstein}
{G^\alpha}_{\beta}+\Lambda \, {g^\alpha}_\beta
= \mu \, \, {T^\alpha}_\beta, 
\end{equation}
where ${G^\alpha}_\beta$ is the Einstein tensor, $\Lambda$ is the cosmological
constant, and $\mu$ is the Einstein gravitational constant.
When we consider the universe filled with the perfect fluid of the mass density $\rho$
and the pressure $p$, we are able to use the stress-energy tensor
${T^\alpha}_\beta$ of the perfect fluid given by 
\[
{T^\alpha}_\beta:=\mbox{diag} (-\rho c^2,p,\dots,p).
\]
When the cosmological term $\Lambda {g^\alpha}_\beta$ is transposed to
right-hand-side in \eqref{Einstein}, the term is regarded as a part of the
stress-energy tensor.
Then the term $-\Lambda {g^\alpha}_\beta$ is rewritten as 
\[
-\Lambda\, {g^\alpha}_\beta
=\mu \, \mbox{diag} (-\widetilde{\rho} c^2,\widetilde{p},\dots,
\widetilde{p}),
\ \ 
\widetilde{\rho}:=\frac{\Lambda}{\mu c^2}, 
\ \ 
\widetilde{p}:=-\frac{\Lambda}{\mu},
\]
{and} 
the cosmological constant $\Lambda>0$ is regarded 
as the energy which has positive density and negative pressure.
Thus, we regard the cosmological constant $\Lambda$ as ``the dark energy." 
The study of roles of the cosmological constant 
and the spatial variance is important to describe the history of the universe, 
especially, the inflation and the 
accelerating expansion of the universe 
(see e.g., 
\cite{Guth-1981-PhysRevD}, 
\cite{Kazanas-1980-AstrophysicalJ},
\cite{Perlmutter-etal-1999-AstrophysicalJ},
\cite{Riess-Schmidt-etal-1998-AstronomicalJ},
\cite{Sato-1981-MonNotRAstrSoc}, 
\cite{Starobinsky-1980-PhysicsLettersB}).  
One of the solutions of the equation \eqref{Einstein} is the de Sitter
spacetime, 
and its line-element with 
the spatial zero-curvature is given by 
\beq
\label{deSitter-Metric}
ds^2
=
g_{\alpha\beta} dx^\alpha dx^\beta
=
-c^2dt^2+e^{2Ht}\delta_{ij} dx^idx^j,
\eeq
where {$t:=x^0$}, $H$ is the Hubble constant, and $\delta_{ij}$ denotes 
the Kronecker delta.
The relation between the cosmological constant $\Lambda$ and the Hubble
constant $H$ is given by $H=\sqrt{2\Lambda/(n^2-n)}$.
We can write 
$(g_{\alpha\beta})_{0\le \alpha,\beta\le n}={\rm diag}(-c^2, e^{2Ht}, \cdots, e^{2Ht})$.

Next, we introduce the semilinear Klein-Gordon equation (KGE) in curved
spacetimes.
KGE presents the equation of motion for the massive scalar field.
Then, to derive the equation of motion described by a real-valued function
$\phi$ with the mass $m$ and the potential $V$, we consider the Lagrangian
density $\mathcal{L}$ defined by
\begin{align}
\mathcal{L}
:=
-\frac{\sqrt{-g}}{2}
\left\{
g^{\alpha\beta}(\nabla_\alpha\phi)(\nabla_\beta \phi)
+
\frac{c^2m^2}{\hbar^2} \phi^2
+2V(\phi)
\right\},
\label{eq:Lagrangian}
\end{align}
where $\nabla_\mu$ is the covariant derivative associated with $g_{\mu\nu}$
and $\hbar$ is the Planck constant.
If the spacetime is flat, 
the Lagrangian density $\mathcal{L}$ is
consistent with the well-known Lagrangian density of KGE (see e.g., \cite{Wald}).
The Euler-Lagrange equation of the action 
$\int_{\br^{1+n}} \mathcal{L} \, 
%\sqrt{-g} 
dtdx$ gives KGE in curved spacetime
\[
-\frac{1}{ \sqrt{-g} } \, \partial_\alpha (\sqrt{-g} g^{\alpha\beta}\partial_\beta \phi)
+ \frac{m^2 c^2}{\hbar^2}\phi+V'(\phi)=0.
\]
In the de Sitter spacetime, KGE is rewritten as 
\beq
\label{KG-deSitter}
\partial_t^2\phi+nH\partial_t\phi-c^2e^{-2Ht}
\delta^{ij}\partial_i\partial_j \phi
+\frac{m^2c^4}{\hbar^2}\phi+c^2V'(\phi)=0,
\eeq
where $V'$ denotes the derivative $V'(r)=dV(r)/dr$ for $r\in \br$ 
(see, e.g., 
\cite{Baskin-2012-AHP, 
Nakamura-2014-JMAA,
Yagdjian-Galstian-2009-CMP}).
Multiplying $\partial_t \phi$ to the both sides of \eqref{KG-deSitter}, 
we have the divergence formula 
\[
\partial_t e^0+\partial_j e^j+e_\ast =0,
\]
where $e^\alpha$, $0\le \alpha\le n$, and $e_\ast$ mean the energy density
and the external energy, respectively.
They are defined by 
\begin{eqnarray*}
e^0&:=&
\frac{1}{2}
\left\{
(\partial_t \phi)^2+c^2 e^{-2Ht}\delta^{ij}(\partial_i\phi)(\partial_j\phi)
+\frac{m^2c^4}{\hbar^2} \phi^2+2c^2V(\phi)
\right\}, 
\\
e^j&:=&-c^2e^{-2Ht} \delta^{ij}(\partial_i \phi)(\partial_t\phi),
\\
e_\ast&:=&
H\{n(\partial_t\phi)^2
+ c^2e^{-2Ht}\delta^{ij}(\partial_i\phi)(\partial_j\phi)\}.
\end{eqnarray*}
We define the total energy by 
\begin{align}
  E(t):=\int_\brn e^0(t,x) dx,
  \label{eq:totalE}
\end{align}
then we have the energy estimate 
\beq
\label{E-Int}
E(t)+\int_0^t \int_{\brn} e_\ast (s,x) dx\,ds=E(0)
\eeq
for $t\in [0,\infty)$.
Let us consider the defocusing potential ($V\ge0$), 
by which $E$ is a positive-valued function for any non-trivial solution $\phi$.
When the space is expanding ($H>0$), 
the estimate \eqref{E-Int} shows the dissipative property $E(t)< E(0)$ for $t>0$.
On the other hand, 
when the space is contracting ($H<0$),
\eqref{E-Int} shows the anti-dissipative property $E(t)> E(0)$ for $t>0$.

We introduce the known theoretical results 
on KGE \eqref{KG-deSitter}.
Let us consider the scalar potential $V$ of power type and its derivative $V'$ given by 
\begin{equation}
\label{Def-V}
V(\phi):=\frac{\lambda}{p+1} |\phi|^{p+1},
\ \ \ \ 
V'(\phi)=\lambda |\phi|^{p-1}\phi
\end{equation}
for $\lambda\in \br$ and $1<p<\infty$.
We consider the Cauchy problem 
\begin{equation}
\label{Cauchy}
\left\{
\begin{array}{l}
\left(
\partial_t^2\phi+nH\partial_t\phi-c^2e^{-2Ht}
\delta^{ij}\partial_i\partial_j \phi+m^2c^4\phi/\hbar^2
+c^2
\lambda |\phi|^{p-1}\phi
\right)(t,x)
=0\\
\hfill \ \ \ \ \mbox{for}\ \ (t,x)\in [0,T)\times\br^n \\
\phi(0,\cdot)=\phi_0(\cdot)\in H^1(\br^n),
\ \ 
\partial_t\phi(0,\cdot)=\phi_1(\cdot)\in L^2(\br^n),
\end{array}
\right.
\end{equation}
where $\phi_0$, $\phi_1$ are given initial data, 
$H^1(\mathbb{R}^n)$ denotes the Sobolev space 
and $L^2(\mathbb{R}^n)$ denotes the Lebesgue space.
On the Cauchy problem \eqref{Cauchy} for the positive Hubble constant $H>0$, 
the following theoretical results have been shown.
The first result shows the existence of the time-local solutions for any initial data.
The second result shows that the local solutions are time-global solutions 
if the initial data are sufficiently small.
The third result shows the local solutions are global for any data 
if the semilinear term is defocusing.

\begin{theorem}
\label{Theorem}
(\cite[Theorems 1.2 and 1.7]{Nakamura-2014-JMAA})  
Let $H>0$. 
Assume $m>nH/2$.
Let $p$ satisfy 
\beq
\label{Ineq-BoundP}
1\le p 
\left\{
\begin{array}{ll}
<   \infty &\mbox{if}\ \ n=1,2 \\
\le 1+\frac{2}{n-2} &\mbox{if}\ \ n\ge3.
\end{array}
\right.
\eeq
Then we have the following results.

(1) 
For any $\phi_0$ and $\phi_1$, there exists 
$T=T(\|\phi_0\|_{H^1(\brn)}+\|\phi_1\|_{L^2(\brn)} )>0$ 
such that \eqref{Cauchy} has a unique solution 
$\phi$ in  $C([0,T),H^1(\brn))\cap C^1([0,T),L^2(\brn))$. 

(2) 
Let $n\le 4$.
If $\|\phi_0\|_{H^1(\brn)}+\|\phi_1\|_{L^2(\brn)}$ is sufficiently small, and $1+4/n\le p$, 
then \eqref{Cauchy} has a unique solution $\phi$ in $C([0,\infty),H^1(\brn))\cap C^1([0,\infty),L^2(\brn))$.

(3) 
If $\lambda\ge0$, then \eqref{Cauchy} has a unique global solution 
$\phi$ in $C([0,\infty),H^1(\br^n))\cap C^1([0,\infty),L^2(\br^n))$ 
for any data $\phi_0\in H^1(\mathbb{R}^n)$ and $\phi_1\in L^2(\mathbb{R}^n)$. 
\end{theorem}

%%%%%%%%%%%%%%%%%%%%%%%%%
% Known results
%%%%%%%%%%%%%%%%%%%%%%%%%

We refer to the corresponding known results on \eqref{Cauchy}.
D'Ancona and Giuseppe have shown in 
\cite{D'Ancona-1995-MathZ} and 
\cite{D'Ancona-Giuseppe-2001-MathNachr} 
global classical solutions for $(\partial_t^2-a(t)
\delta^{ij}\partial_i\partial_j
) u+|u|^{p-1}u=0$ with some additional conditions on  $a(t)\ge0$ and $p$ when $n=1,2,3$. 
Yagdjian has shown in 
\cite{Yagdjian-2012-JMAA} 
small global solutions for \eqref{Eq-KGV} with arbitrary $n$  
when the nonlinear term $f$ is of power type $p>1$, 
and the norm of initial data $\|u_0\|_{H^s(\brn)}+\|u_1\|_{H^s(\brn)}$ is sufficiently small for some $s>n/2\ge 1$  
(see also  \cite{Yagdjian-2013-Springer} for the system of the equations). 
Galstian and Yagdjian has extended 
this result to the case of the Riemann metric space for each time slices 
in \cite{Galstian-Yagdjian-2015-NA-RWA}.
In \cite{Nakamura-2014-JMAA}, the energy solutions $(s=1)$  in Theorem \ref{Theorem} have been shown. 
This result has been extended to the case of general Friedmann-Lema\^itre-Robertson-Walker spacetime 
in \cite{Galstian-Yagdjian-2015-NA-TMA}.
Baskin has shown in \cite{Baskin-2012-AHP} small global solution for 
$(g^{\mu\nu}\nabla_\mu \nabla_\nu+\lambda) u+f(u)=0$ 
when $f(u)$ is a type of $|u|^{p-1} u$,  
$p=1+4/(n-1)$, $\lambda>n^2/4$, $(u_0,u_1)\in H^1\oplus L^2$, 
where $g^{\mu\nu}$ gives the asymptotic de Sitter spacetime
(see also \cite{Baskin-2010-ProcCMAANU} for the cases $p=5$ with $n=3$, $p=3$ with $n=4$). 
Blow-up phenomena are considered in \cite{Yagdjian-2009-DCDS}.
See also the references in the summary \cite{Yagdjian-9999-CurvedSpacetime} by Yagdjian.

In Theorem \ref{Theorem}, we have assumed the condition $m>nH/2$ by the
following reason.
Since the first equation in \eqref{Cauchy} has the dissipative
(or anti-dissipative) term $nH\partial_t \phi$, we use the transformation
$u=e^{\kappa t}\phi$ for $\kappa\in \mathbb{R}$ to transform the equation to
the Klein-Gordon equation.
Then the equation is rewritten as 
\begin{multline}
\label{Eq-KGLambda}
\left\{
\partial_t^2+(nH-2\kappa)\partial_t-c^2e^{-2Ht}
\delta^{ij}\partial_i\partial_j 
+\frac{m^2c^4}{\hbar^2}+\kappa(\kappa-nH)
\right\}u \\
+c^2e^{\kappa t}V'(e^{-\kappa t}u)=0,
\end{multline}
and we obtain the equation  
\beq
\label{Eq-KGV}
(\partial_t^2-c^2e^{-2Ht} 
\delta^{ij}\partial_i\partial_j+M^2)u(t,x)
+c^2e^{nHt/2}V'(e^{-nHt/2}u(t,x))=0
\eeq
when $\kappa=nH/2$, $M^2:=c^4m^2/\hbar^2-n^2H^2/4$.
We assume $m$ is large such that $M^2>0$ 
(see \cite{Yagdjian-Galstian-2009-CMP}). 
Here, we expect some dissipative effects when $\kappa<nH/2$ in \eqref{Eq-KGLambda}.
However, $\kappa\le nH/2$ is needed for the energy estimates for \eqref{Eq-KGV}, 
and $\kappa\ge nH/2$ is needed for the contraction argument to show the existence of the solutions of \eqref{Eq-KGV}.
So that, we assume $\kappa=nH/2$, 
by which we lose the dissipative term $(nH-2\kappa)\partial_tu$ in \eqref{Eq-KGV}.
Since the first equation in \eqref{Cauchy} has the term 
$e^{-2Ht}\delta^{ij}\partial_i\partial_j \phi$
with the non-constant coefficient, it is not easy to obtain the critical
theoretical results so far.
Instead, in this paper, we carry out some numerical simulations, and show the
detailed behaviors to clarify the dissipative effect of the spatial
expansion.

\newsection{Hamiltonian formulation of KGE}

Let us consider the Hamiltonian density for the Lagrangian density
$\mathcal{L}$.
The Lagrangian density of KGE 
\eqref{eq:Lagrangian} in the de Sitter spacetime becomes
\begin{align}
  \mathcal{L}
  &= -\frac{c}{2}e^{nHt}\left\{
  - \frac{1}{c^2}(\partial_t\phi)^2
  + e^{-2Ht}\delta^{ij}(\partial_i\phi)(\partial_j\phi)
  + \frac{c^2m^2}{\hbar^2}\phi^2
  + 2V(\phi)
  \right\}.
\end{align}
Then, we define the momentum $\psi$ by 
\[
\psi:=\frac{\partial \mathcal{L} }{\partial (\partial_t \phi)}
=\frac{e^{nHt}}{c}\, \partial_t \phi,
\]
the Hamiltonian density $\mathcal{H}$ by the Legendre transformation
\begin{eqnarray*}
  \mathcal{H}
  &:=&\psi (\partial_t\phi)-\mathcal{L}\\
  &=&
  \frac{ ce^{nHt} }{2}
\left\{
e^{-2nHt}\psi^2
+e^{-2Ht}
\delta^{ij}(\partial_i\phi)(\partial_j\phi)
+
\frac{c^2m^2}{\hbar^2} \phi^2+2V(\phi)
\right\},
\end{eqnarray*}
and the Hamiltonian $H_{C}$ by 
\begin{align}
  H_{C}(t):= \int_{\br^n} \mathcal{H}(t,x) dx
  \label{eq:totalH}
\end{align}
for $t\in \br$.
To investigate the properties of $H_C$, we denote the kinetic term,
the diffusion term, the mass term and the nonlinear term by the integration of
$(\partial_t\phi)^2/2$,
$c^2e^{-2Ht}\delta^{ij}(\partial_i\phi)(\partial_j\phi)/2$, 
$m^2c^4\phi^2/(2\hbar^2)$ and $c^2V(\phi)$.
Namely, 
\begin{align}
\begin{array}{lll}
  \mbox{the kinetic term} &: &
  K(t):=\displaystyle \int_{\br^n} \frac{1}{2}
  (\partial_t\phi(t,x))^2 dx, \\
  \mbox{the diffusion term} &: &
  D(t):=\displaystyle \int_{\br^n}
  \frac{1}{2}c^2e^{-2Ht}\delta^{ij}(\partial_i \phi(t,x))
  (\partial_j \phi(t,x))dx, \\
  \mbox{the mass term} &: &
  M(t):=\displaystyle \int_{\br^n}  \frac{m^2c^4}{2\hbar^2}
  \phi(t,x)^2dx,
  \\
  \mbox{the nonlinear term} &: &
  N(t):=\displaystyle \int_{\br^n} c^2V(\phi(t,x)) dx
  =\displaystyle \int_{\br^n} \frac{\lambda c^2}{p+1}|\phi|^{p+1} dx
\end{array}
\label{eq:componentsE}
\end{align}
for $t\in \br$.
With \eqref{eq:componentsE}, the total energy \eqref{eq:totalE} can be
expressed as
\begin{align}
  E(t)
  = K(t) + D(t) + M(t) + N(t).
\end{align}
Since the relation between $\mathcal{H}$ and the energy density $e_0$ becomes
\begin{align}
  \mathcal{H} = \frac{e^{nHt}}{c}e_0,
\end{align}
$H_C$ can be expressed as
\begin{align*}
  H_C(t) = \frac{e^{nHt}}{c}(K(t) + D(t) + M(t) + N(t)).
\end{align*}
If the spacetime is flat ($H=0$) and the speed of light $c$ is set as 1,
then $H_C$ is consistent with $E$.
The Hamiltonian formulation of \eqref{KG-deSitter} is given by the canonical
equations
\begin{align}
  \begin{array}{ll}
  \partial_t\phi
  &:=\displaystyle{\frac{\delta\mathcal{H}}{\delta \psi}
  = ce^{-nHt}\psi},\\
  \partial_t\psi
  &:=\displaystyle{
  -\frac{\delta\mathcal{H}}{\delta \phi}
  = ce^{(n-2)Ht}\delta^{ij}(\partial_i\partial_j\phi)
  - \frac{c^3m^2}{\hbar^2}e^{nHt}\phi
  - ce^{nHt}V'(\phi)}.
  \end{array}
  \label{eq:HamitonEvoKGE}
\end{align}
Since $\mathcal{H}$ satisfies the differential equation
\[
\partial_t \mathcal{H}
=
nH\mathcal{H}
+c^2e^{-2Ht}\partial_i\{\psi \delta^{ij}(\partial_j\phi)\}
-\mathcal{H}_\ast,
\]
where we have put 
\[
\mathcal{H}_\ast
=
H\{cne^{-nHt}\psi^2
+
ce^{(n-2)Ht} \delta^{ij}(\partial_i\phi)(\partial_j\phi)\},
\]
$H_C$ satisfies the equation
\begin{align}
  \partial_t H_C(t)
  &=
  nHH_C(t)- \int_{\mathbb{R}^n}\mathcal{H}_\ast\,dx,
  \label{HCt-Int}
\end{align}
and by using \eqref{eq:componentsE}, \eqref{HCt-Int} can be expressed as
\begin{align}
  \partial_t H_C(t)
  &=
  \frac{H}{c}e^{nHt}\{-nK(t) + (n-2)D(t) + nM(t) + nN(t)\}.
  \label{HCt-Int2}
\end{align}
The equation \eqref{HCt-Int} (or \eqref{HCt-Int2}) indicates $H_C$ is not
conserved in time evolution 
if the Hubble constant $H$ is not zero.
Therefore, we define $\tilde{H}_C$ as
\begin{align}
  \tilde{H}_C(t)
  := H_C(t) - \int^t_0 \partial_t H_C(s)ds,
  \label{eq:modTotalH}
\end{align}
which is a constant value in time evolution independent of the value of $H$.

The behaviors of $E$ and $H_C$ show the properties of the effects 
of the spatial expansion and contraction by the Hubble constant $H$.
Namely, we are able to understand the properties of the dark energy, 
which is equivalent to the cosmological constant 
mathematically, through the asymptotic behaviors of  $\phi$, $E$ and $H_C$.
However, it is not easy to obtain the detailed behaviors of them by theoretical
methods since the waves of $\phi$ would be oscillating and interfered by the
semilinear term $V(\phi)$, and \eqref{KG-deSitter} is the equation with the
variable coefficient.
Thus, we perform the numerical simulations on $\phi$, $E$ and $H_C$ to study
the detailed dissipative effects of the spatial expansion.

%\newsection{Discrete KGE with SPS}
\newsection{\textcolor{black}{Discrete KGE}}

To investigate KGE by \textcolor{black}{computational} analysis, we need the high precise and accurate
numerical data.
%This is because that if we get the poor numerical data, then we cannot
%investigate the properties of KGE.
Then, we use SPS to get the high precise and accurate numerical results.
With this scheme, the structures of the equations in the continuous case are
preserved in the discrete case.
In this paper, we use this scheme to make the discrete Hamiltonian
formulation.
The discrete values of the variable $u$ are defined as 
$u{}^{(\ell)}_{(k)}
:=u(\ell\Delta t, k\Delta x)$, where $\ell$ is the time index, $\Delta t$ is the time
mesh width, $k$ is the spatial grid index, 
and $\Delta x$ is the spatial mesh width.
The details of this are in Ref.\cite{Furihata-Matsuo-2010}.

\textcolor{black}{
  For comparison with SPS, we use the Crank-Nicolson scheme (CNS) and the
  fourth-order Runge-Kutta scheme (RKS) which are widely
  used for calculating partial differential equations numerically.
}

\subsection{\textcolor{black}{SPS}}

By using SPS, we can get a set of discrete evolution equations of KGE.
We give a discrete Hamiltonian density $\mathcal{H}^{(\ell)}_{(k)}$ as
\begin{align}
  \mathcal{H}^{(\ell)}_{(k)}
  &= \frac{ce^{nHt_\ell}}{2}\biggl\{
  e^{-2nHt_\ell}(\psi^{(\ell)}_{(k)})^2
  + e^{-2Ht_\ell}\delta^{ij}
  (\widehat{\delta}^{\langle1\rangle}_i\phi^{(\ell)}_{(k)})
  (\widehat{\delta}^{\langle1\rangle}_j\phi^{(\ell)}_{(k)})
  + \frac{c^2m^2}{2\hbar^2}(\phi^{(\ell)}_{(k)})^2
  \nonumber\\
  &\quad
  + \frac{2\lambda}{p+1}|\phi^{(\ell)}_{(k)}|^{p+1}\biggr\},
\end{align}
where $t_\ell$ presents the time of the $\ell$th step and 
$\widehat{\delta}^{\langle1\rangle}_iu^{(\ell)}_{(k)}$ is the centered space
operator defined by
\begin{align*}
  \widehat{\delta}^{\langle1\rangle}_iu^{(\ell)}_{(k)}
  :=
  \frac{u^{(\ell)}_{(k+1)} - 2u^{(\ell)}_{(k)} + u^{(\ell)}_{(k-1)}}{2\Delta x^i}.
\end{align*}
For $\mathcal{H}^{(\ell)}_{(k)}$, $\phi^{(\ell)}_{(k)}$ and 
$\psi^{(\ell)}_{(k)}$, 
we define the values 
\[
\widehat{\delta}{\mathcal{H}}/\bigl(\widehat{\delta}
(\phi^{(\ell+1)}_{(k)}, \phi^{(\ell)}_{(k)})\bigr)
\ \ 
\mbox{and}
\ \  
\widehat{\delta} \mathcal{H}/\bigl(\widehat{\delta}(\psi^{(\ell+1)}_{(k)},
\psi^{(\ell)}_{(k)})\bigr)
\]
by the equation
\begin{align}
  \mathcal{H}^{(\ell+1)}_{(k)} - \mathcal{H}^{(\ell)}_{(k)}
  &=
  \frac{\widehat{\delta} \mathcal{H}}{\widehat{\delta}(\phi^{(\ell+1)}_{(k)},
    \phi^{(\ell)}_{(k)})}(\phi^{(\ell+1)}_{(k)}-\phi^{(\ell)}_{(k)})
  + \frac{\widehat{\delta} \mathcal{H}}{\widehat{\delta}(\psi^{(\ell+1)}_{(k)},
    \psi^{(\ell)}_{(k)})}(\psi^{(\ell+1)}_{(k)}-\psi^{(\ell)}_{(k)}).
\end{align}
Then, a set of discrete KGE with SPS becomes
\begin{align}
  \begin{array}{ll}
  \displaystyle{\frac{\phi^{(\ell+1)}_{(k)} - \phi^{(\ell)}_{(k)}}{\Delta t}}
  &=\displaystyle{
    \frac{\widehat{\delta} \mathcal{H}}{\widehat{\delta}(\psi^{(\ell+1)}_{(k)},
      \psi^{(\ell)}_{(k)})}}\\
  &=\displaystyle{
    \frac{1}{4}c(e^{-nHt_{\ell+1}} + e^{-nHt_\ell})
    (\psi^{(\ell+1)}_{(k)} + \psi^{(\ell)}_{(k)})},
  \\
  \displaystyle{\frac{\psi^{(\ell+1)}_{(k)} - \psi^{(\ell)}_{(k)}}{\Delta t}}
  &=\displaystyle{
    -\frac{\widehat{\delta} \mathcal{H}}{\widehat{\delta}(\phi^{(\ell+1)}_{(k)},
    \phi^{(\ell)}_{(k)})}}\\
  &=
  \displaystyle{
    \frac{1}{4}c(e^{(n-2)Ht_{\ell+1}}+e^{(n-2)Ht_\ell})
    \delta^{ij}\widehat{\delta}^{\langle1\rangle}_j
    \widehat{\delta}^{\langle1\rangle}_i(\phi^{(\ell+1)}_{(k)} + \phi^{(\ell)}_{(k)})
  }\\
  &\quad
  \displaystyle{
    - \frac{c^3m^2}{4\hbar^2}(e^{nHt_{\ell+1}} + e^{nHt_\ell})
    (\phi^{(\ell+1)}_{(k)} + \phi^{(\ell)}_{(k)})}\\
  &\quad
  \displaystyle{
    - \frac{\lambda c}{2(p+1)}(e^{nHt_{\ell+1}} +e^{nHt_\ell})
  \frac{|\phi^{(\ell+1)}_{(k)}|^{p+1}
    - |\phi^{(\ell)}_{(k)}|^{p+1}}{\phi^{(\ell+1)}_{(k)}- \phi^{(\ell)}_{(k)}}.
  }
  \end{array}
  \label{eq:DiscreteHamitonEvoKGE}
\end{align}
The discrete evolution equations \eqref{eq:DiscreteHamitonEvoKGE}
are corresponding to \eqref{eq:HamitonEvoKGE} for the continuous case.
The nonlinear term of the last term in \eqref{eq:DiscreteHamitonEvoKGE} is
expressed as
\begin{align}
  &\frac{|\phi^{(\ell+1)}_{(k)}|^{p+1} - |\phi^{(\ell)}_{(k)}|^{p+1}}{\phi^{(\ell+1)}_{(k)}- \phi^{(\ell)}_{(k)}}\nonumber\\
  &= \{|\phi^{(\ell+1)}_{(k)}|^{p}
  + |\phi^{(\ell+1)}_{(k)}|^{p-1}|\phi^{(\ell)}_{(k)}|
  + \cdots
  + |\phi^{(\ell+1)}_{(k)}||\phi^{(\ell)}_{(k)}|^{p-1}
  + |\phi^{(\ell)}_{(k)}|^{p}\}
  \frac{|\phi^{(\ell+1)}_{(k)}| - |\phi^{(\ell)}_{(k)}|}{
    \phi^{(\ell+1)}_{(k)} - \phi^{(\ell)}_{(k)}}
\end{align}
{when $p$ is a natural number.}
We refer to the expression in the nonlinear Schr\"odinger equations in
Ref.\cite{Furihata-Matsuo-2010} to make above relation.
With \eqref{eq:DiscreteHamitonEvoKGE}, the time difference
of $\mathcal{H}^{(\ell)}_{(k)}$ is calculated as
\begin{align}
  &\frac{\mathcal{H}^{(\ell+1)}_{(k)} - \mathcal{H}^{(\ell)}_{(k)}}{\Delta t}
  \nonumber\\
  &= \frac{c}{4}\{(\psi^{(\ell+1)}_{(k)})^2 + (\psi^{(\ell)}_{(k)})^2\}
  \frac{e^{-nHt_{\ell+1}}-e^{-nHt_\ell}}{\Delta t}
  \nonumber\\
  &\quad
  + \frac{c}{4}(e^{-nHt_{\ell+1}} + e^{-nHt_\ell})
  (\psi^{(\ell+1)}_{(k)} + \psi^{(\ell)}_{(k)})
  \frac{\psi^{(\ell+1)}_{(k)} - \psi^{(\ell)}_{(k)}}{\Delta t}
  \nonumber\\
  &\quad
  + \frac{c}{4}\delta^{ij}
  \{(\widehat{\delta}^{\langle1\rangle}_i\phi^{(\ell+1)}_{(k)})
  (\widehat{\delta}^{\langle1\rangle}_j\phi^{(\ell+1)}_{(k)})
  + (\widehat{\delta}^{\langle1\rangle}_i\phi^{(\ell)}_{(k)})
  (\widehat{\delta}^{\langle1\rangle}_j\phi^{(\ell)}_{(k)})\}
  \frac{e^{(n-2)Ht_{\ell+1}}-e^{(n-2)Ht_\ell}}{\Delta t}
  \nonumber\\
  &\quad
  + \frac{c}{4}(e^{(n-2)Ht_{\ell+1}}+e^{(n-2)Ht_\ell})
  \delta^{ij}\widehat{\delta}^{\langle1\rangle}_i
  (\phi^{(\ell+1)}_{(k)} + \phi^{(\ell)}_{(k)})
  \widehat{\delta}^{\langle1\rangle}_j
  \frac{\phi^{(\ell+1)}_{(k)} - \phi^{(\ell)}_{(k)}}{\Delta t}
  \nonumber\\
  &\quad
  + \frac{c^3m^2}{4\hbar^2}\{(\phi^{(\ell+1)}_{(k)})^2
  + (\phi^{(\ell)}_{(k)})^2\}
  \frac{e^{nHt_{\ell+1}} - e^{nHt_\ell}}{\Delta t}
  \nonumber\\
  &\quad
  + \frac{c^3m^2}{4\hbar^2}(e^{nHt_{\ell+1}} + e^{nHt_\ell})
  (\phi^{(\ell+1)}_{(k)} + \phi^{(\ell)}_{(k)})
  \frac{\phi^{(\ell+1)}_{(k)} - \phi^{(\ell)}_{(k)}}{\Delta t}
  \nonumber\\
  &\quad
  + \frac{c\lambda}{2(p+1)}
  (|\phi^{(\ell+1)}_{(k)}|^{p+1}+|\phi^{(\ell)}_{(k)}|^{p+1})
  \frac{e^{nHt_{\ell+1}}-e^{nHt_\ell}}{\Delta t}
  \nonumber\\
  &\quad
  + \frac{c\lambda}{2(p+1)}(e^{nHt_{\ell+1}}+e^{nHt_\ell})
  \frac{|\phi^{(\ell+1)}_{(k)}|^{p+1}
    - |\phi^{(\ell)}_{(k)}|^{p+1}}{\phi^{(\ell+1)}_{(k)} - \phi^{(\ell)}_{(k)}}
  \frac{\phi^{(\ell+1)}_{(k)} - \phi^{(\ell)}_{(k)}}{\Delta t}
  \nonumber
\end{align}
\begin{align}
  &=
  \frac{c}{4}\{(\psi^{(\ell+1)}_{(k)})^2 + (\psi^{(\ell)}_{(k)})^2\}
  \frac{e^{-nHt_{\ell+1}}-e^{-nHt_\ell}}{\Delta t}
  \nonumber\\
  &\quad
  + \frac{c^2}{16}
  (e^{-nHt_{\ell+1}} + e^{-nHt_\ell})
  (e^{(n-2)Ht_{\ell+1}} + e^{(n-2)Ht_\ell})
  \nonumber\\
  &\hspace{15mm}\quad\cdot
  \delta^{ij}\widehat{\delta}^{\langle1\rangle}_i\{
  (\psi^{(\ell+1)}_{(k)} + \psi^{(\ell)}_{(k)})
  \widehat{\delta}^{\langle1\rangle}_j
  (\phi^{(\ell+1)}_{(k)} + \phi^{(\ell)}_{(k)})\}\quad
  \nonumber\\
  &\quad
  + \frac{c}{4}\delta^{ij}
  \{(\widehat{\delta}^{\langle1\rangle}_i\phi^{(\ell+1)}_{(k)})
  (\widehat{\delta}^{\langle1\rangle}_j\phi^{(\ell+1)}_{(k)})
  + (\widehat{\delta}^{\langle1\rangle}_i\phi^{(\ell)}_{(k)})
  (\widehat{\delta}^{\langle1\rangle}_j\phi^{(\ell)}_{(k)})\}
  \frac{e^{(n-2)Ht_{\ell+1}}-e^{(n-2)Ht_\ell}}{\Delta t}
  \nonumber\\
  &\quad
  + \frac{c^3m^2}{4\hbar^2}\{(\phi^{(\ell+1)}_{(k)})^2
  + (\phi^{(\ell)}_{(k)})^2\}
  \frac{e^{nHt_{\ell+1}} - e^{nHt_\ell}}{\Delta t}
  \nonumber\\
  &\quad
  + \frac{c\lambda}{2(p+1)}
  (|\phi^{(\ell+1)}_{(k)}|^{p+1}+|\phi^{(\ell)}_{(k)}|^{p+1})
  \frac{e^{nHt_{\ell+1}}-e^{nHt_\ell}}{\Delta t}
  \nonumber
\end{align}
\begin{align}
  &=
    - \frac{cnH}{4}\{(\psi^{(\ell+1)}_{(k)})^2 + (\psi^{(\ell)}_{(k)})^2\}e^{-nHt_\ell}
  \nonumber\\
  &\quad
    + \widehat{\delta}^{\langle1\rangle}_i\biggl\{\frac{c^2}{16}
    (e^{-nHt_{\ell+1}} + e^{-nHt_\ell})
    (e^{(n-2)Ht_{\ell+1}} + e^{(n-2)Ht_\ell})
  \nonumber\\
  &\hspace{15mm}\quad\cdot
    \{
    (\psi^{(\ell+1)}_{(k)} + \psi^{(\ell)}_{(k)})
    \delta^{ij}\widehat{\delta}^{\langle1\rangle}_j
    (\phi^{(\ell+1)}_{(k)} + \phi^{(\ell)}_{(k)})\}\biggr\}
  \nonumber\\
  &\quad
    + \frac{c(n-2)H}{4}\delta^{ij}
    \{(\widehat{\delta}^{\langle1\rangle}_i\phi^{(\ell+1)}_{(k)})
    (\widehat{\delta}^{\langle1\rangle}_j\phi^{(\ell+1)}_{(k)})
    + (\widehat{\delta}^{\langle1\rangle}_i\phi^{(\ell)}_{(k)})
    (\widehat{\delta}^{\langle1\rangle}_j\phi^{(\ell)}_{(k)})\}
    e^{(n-2)Ht_\ell}
  \nonumber\\
  &\quad
    + \frac{c^3m^2nH}{4\hbar^2}\{(\phi^{(\ell+1)}_{(k)})^2
    + (\phi^{(\ell)}_{(k)})^2\}e^{nHt_\ell}
  \nonumber\\
  &\quad
    + \frac{c\lambda nH}{2(p+1)}
    (|\phi^{(\ell+1)}_{(k)}|^{p+1}+|\phi^{(\ell)}_{(k)}|^{p+1})e^{nHt_\ell}
    + O(\Delta t),
  \label{eq:timeDiffH}
\end{align}
where we use the relation of $t_{\ell+1}=t_\ell+\Delta t$.
We define the discrete Hamiltonian $H_C{}^{(\ell)}$ by
\begin{align}
    H_C{}^{(\ell)}
    := \sum_{k{ \in \mathbb{Z} } }\mathcal{H}^{(\ell)}_{(k)}
    \prod_{ {1\le i\le n} }
    \Delta x^i,
    \label{eq:definedHC}
  \end{align}
  then \eqref{eq:timeDiffH} is calculated as
\begin{align}
  \frac{H_C{}^{(\ell+1)} - H_C{}^{(\ell)}}{\Delta t}
  &=
  \frac{H}{2c}e^{nHt_\ell}\{
  - n(K^{(\ell+1)} + K^{(\ell)})
  + (n-2)(D^{(\ell+1)} + D^{(\ell)})
  \nonumber\\
  &\quad
  + n(M^{(\ell+1)} + M^{(\ell)})
  + n(N^{(\ell+1)} + N^{(\ell)})\}
  + O(\Delta t),
  \label{eq:DiscreteE}\end{align}
where $K^{(\ell)}$, $D^{(\ell)}$, $M^{(\ell)}$ and $N^{(\ell)}$ are defined by
\begin{align}
  \begin{array}{l}
  K{}^{(\ell)}
  := \displaystyle{\sum_{k{ \in \mathbb{Z} } } 
  \frac{1}{2}c^2e^{-2nHt_\ell}(\psi^{(\ell)}_{(k)})^2
    \prod_{{1\le i\le n} } \Delta x^i},\\
  D{}^{(\ell)}
  := \displaystyle{\sum_{k{ \in \mathbb{Z} } } 
  \frac{1}{2}c^2e^{-2Ht_\ell}\delta^{ij}
  (\widehat{\delta}^{\langle1\rangle}_i\phi^{(\ell)}_{(k)})
  (\widehat{\delta}^{\langle1\rangle}_j\phi^{(\ell)}_{(k)})
    \prod_{{1\le i\le n} } \Delta x^i},\\
  M{}^{(\ell)}
  := \displaystyle{\sum_{k{ \in \mathbb{Z} } } 
  \frac{m^2c^4}{2\hbar^2}
    (\phi^{(\ell)}_{(k)})^2\prod_{{1\le i\le n}} \Delta x^i},\\
  N{}^{(\ell)}
  := \displaystyle{\sum_{k{ \in \mathbb{Z} } } 
  c^2V(\phi^{(\ell)}_{(k)})\prod_{{1\le i\le n} } \Delta x^i}
  = \displaystyle{\sum_{k{ \in \mathbb{Z} } } 
  \frac{\lambda c^2}{p+1}|\phi^{(\ell)}_{(k)}|^{p+1}
  \prod_{ {1\le i\le n} } \Delta x^i}.\\
  \end{array}
  \label{eq:definedDiscreteTerms}
\end{align}
$K^{(\ell)}$, $D^{(\ell)}$, $M^{(\ell)}$ and $N^{(\ell)}$ are discrete values corresponding
to $K(t)$, $D(t)$, $M(t)$ and $N(t)$ defined by \eqref{eq:componentsE},
respectively.
We set
\begin{align}
    \tilde{H}_C{}^{ ({\ell}) }
  &:=
    H_C{}^{({\ell})}
    - \sum_{ {0\le i\le \ell-1} }
    \frac{H}{2c}e^{nHt_\ell}\{
    - n(K^{({i}+1)} + K^{({i})})
    + (n-2)(D^{({i}+1)} + D^{({i})})
  \nonumber\\
  &\quad
    + n(M^{({i}+1)} + M^{({i})})
    + n(N^{({i}+1)} + N^{({i})})\},
  \label{eq:definedHCtilde}
\end{align}
then %{\underline{,}}
$\tilde{H}_C{}^{(n)}$ is a discretized value of $\tilde{H}_C$ defined by
\eqref{eq:modTotalH}.
This value is a guideline to perform precise numerical simulations in the
cases of $H\neq 0$.

{\color{black}

  \subsection{CNS and RKS}
  CNS and RKS are frequently used to perform partial differential equations.
  The discrete equations of \eqref{eq:HamitonEvoKGE} with CNS are given by
  %% \begin{align}
  %%   \begin{array}{ll}
  %%     \displaystyle{\frac{\phi^{(\ell+1)}_{(k)} - \phi^{(\ell)}_{(k)}}{\Delta t}}
  %%     &=\displaystyle{
  %%       \frac{1}{2}c(e^{-nHt_{\ell+1}}\psi^{(\ell+1)}_{(k)}
  %%       + e^{-nHt_\ell}\psi^{(\ell)}_{(k)})},
  %%     \\
  %%     \displaystyle{\frac{\psi^{(\ell+1)}_{(k)} - \psi^{(\ell)}_{(k)}}{\Delta t}}
  %%     &=
  %%     \displaystyle{
  %%       \frac{1}{2}c(e^{(n-2)Ht_{\ell+1}}
  %%       \delta^{ij}\widehat{\delta}^{\langle1\rangle}_j
  %%       \widehat{\delta}^{\langle1\rangle}_i\phi^{(\ell+1)}_{(k)}
  %%       + e^{(n-2)Ht_\ell}
  %%       \delta^{ij}\widehat{\delta}^{\langle1\rangle}_j
  %%       \widehat{\delta}^{\langle1\rangle}_i\phi^{(\ell)}_{(k)})
  %%     }\\
  %%     &\quad
  %%     \displaystyle{
  %%       - \frac{c^3m^2}{2\hbar^2}(
  %%       e^{nHt_{\ell+1}}\phi^{(\ell+1)}_{(k)}
  %%       + e^{nHt_\ell}\phi^{(\ell)}_{(k)})}\\
  %%     &\quad
  %%     \displaystyle{
  %%       - \frac{\lambda c}{2}
  %%       (e^{nHt_{\ell+1}}|\phi^{(\ell+1)}_{(k)}|^{p-1}\phi^{(\ell+1)}_{(k)}
  %%       + e^{nHt_{\ell}}|\phi^{(\ell)}_{(k)}|^{p-1}\phi^{(\ell)}_{(k)}),
  %%     }
  %%   \end{array}
  %%   \label{eq:CNSEvoKGE}
  %% \end{align}
  \begin{align}
    \begin{array}{ll}
      \displaystyle{\frac{\phi^{(\ell+1)}_{(k)} - \phi^{(\ell)}_{(k)}}{\Delta t}}
      = \Phi\left((\psi^{(\ell+1)}_{(k)}+\psi^{(\ell)}_{(k)})/2,
      (t_{\ell+1}+t_\ell)/2\right),\\
      \displaystyle{\frac{\psi^{(\ell+1)}_{(k)} - \psi^{(\ell)}_{(k)}}{\Delta t}}
      = \Psi\left((\phi^{(\ell+1)}_{(k)}+\phi^{(\ell)}_{(k)})/2,
      (t_{\ell+1}+t_\ell)/2\right),
    \end{array}
    \label{eq:CNSEvoKGE}
  \end{align}
  where the right hand sides of \eqref{eq:CNSEvoKGE} are defined as
  \begin{align}
    \begin{array}{ll}
      \Phi(\psi,t)
      := c e^{-nHt}\psi,\\
      \displaystyle{
        \Psi(\phi,t)
        :=
        ce^{(n-2)Ht}\delta^{ij}\widehat{\delta}^{\langle2\rangle}_{ij}\phi
        - \frac{c^3m^2}{\hbar^2}e^{nHt}\phi
        - \lambda ce^{nHt}|\phi|^{p-1}\phi,
      }\\
    \end{array}
  \end{align}
  and the second-order central difference operator
  $\widehat{\delta}^{\langle2\rangle}_{ij}$ is defined as
  \begin{align}
    \widehat{\delta}^{\langle2\rangle}_{ij}u^{(\ell)}_{(k)}
    :=
  \left\{
  \begin{array}{ll}
    (u^{(\ell)}_{(k+1)} - 2u^{(\ell)}_{(k)} + u^{(\ell)}_{(k-1)})/(\Delta x^i)^2,
    & (i=j)\\
    \widehat{\delta}^{\langle1\rangle}_i
    \widehat{\delta}^{\langle1\rangle}_j
    u^{(\ell)}_{(k)}. & (i\neq j)
  \end{array}
  \right.
  \end{align}

  The discrete equations with RKS are give by
  \begin{align}
    &\begin{array}{l}
      \phi^{(\ell+1)}_{(k)}
      = \phi^{(\ell)}_{(k)} + (h_1+2h_2+2h_3+h_4)\Delta t/6,\\
      \psi^{(\ell+1)}_{(k)}
      = \psi^{(\ell)}_{(k)} + (s_1+2s_2+2s_3+s_4)\Delta t/6,
     \end{array}
  \end{align}
  where 
  \begin{align}
    \begin{array}{l}
      h_1 = \Phi(\psi^{(\ell)}_{(k)},t_\ell),\\
      s_1 = \Psi(\phi^{(\ell)}_{(k)},t_\ell),\\
      h_2 = \Phi(\psi^{(\ell)}_{(k)}+s_1\Delta t/2, t_\ell+\Delta t/2),\\
      s_2 = \Psi(\phi^{(\ell)}_{(k)}+h_1\Delta t/2, t_\ell+\Delta t/2),\\
      h_3 = \Phi(\psi^{(\ell)}_{(k)}+s_2\Delta t/2, t_\ell+\Delta t/2),\\
      s_3 = \Psi(\phi^{(\ell)}_{(k)}+h_2\Delta t/2, t_\ell+\Delta t/2),\\
      h_4 = \Phi(\psi^{(\ell)}_{(k)}+s_3\Delta t, t_\ell+\Delta t),\\
      s_4 = \Psi(\phi^{(\ell)}_{(k)}+h_3\Delta t, t_\ell+\Delta t).
    \end{array}
  \end{align}
}

\newsection{Numerical tests}

In this section, we carry out the numerical simulations on $\phi$, $E$ and $H_C$
(or $\tilde{H}_C$).
Since the physical background of the Einstein equation is for the $1+3$ dimensional spacetime, 
we consider the case $n=3$ in the following.

The numerical settings are below. 
Set $(x^1,x^2,x^3)=(x,y,z)\in \mathbb{R}^3$.

\begin{itemize}
\item Initial condition:
  \begin{align*}
\phi_0=\phi_0(x,y,z)=A\cos(2\pi x),\quad
\phi_1=\phi_1(x,y,z)=2\pi A\sin(2\pi x).
  \end{align*}
\item Numerical domains: $0\leq x\leq 1$, $0\leq t\leq 1000$.
\item Boundary condition: periodic.
\item Grids: $\Delta x=1/200$, and $\Delta t=1/500$.
\item
  Physical settings: mass $m=1$, speed of light $c=1$,
  Planck constant $\hbar=1$.
\item The number of exponent in the nonlinear term:
  $p=2, 3, 4,5,6$.
\end{itemize}
We consider the solution $\phi(t,x,y,z)$ for $(t,x,y,z)\in [0,\infty)\times \mathbb{R}^3$ 
which is a constant along the variables $y$ and $z$ to give simple and reliable simulations. 
More general solutions and data will be considered in the future.
We perform the simulations by changing the values of the amplitude $A$ in the
initial condition,
the coefficient value $\lambda$ of the nonlinear term, and the Hubble constant
$H$.
First, we perform some simulations to study the accuracies with SPS.
For comparison with SPS, we use the Crank-Nicolson scheme (CNS) and the
Runge-Kutta scheme (RKS) which are widely used for calculating partial
differential equations numerically.
Because of the nonlinearity of KGE \eqref{eq:HamitonEvoKGE}, the discretized
equations are calculated implicitly.
Thus, the numerical simulations are performed iteratively.
For SPS and CNS, the iterations are nine times and three times, respectively.
If we calculate more times iteratively for each scheme, the numerical results
become worse in our codes.
For the wave equation 
\[
\partial_t^2 \phi-c^2
\delta^{ij}\partial_i\partial_j 
\phi+c^2V'(\phi)=0
\]
with \eqref{Def-V}, 
namely, $H=0$, $m=0$ in \eqref{KG-deSitter} with \eqref{Def-V},
the scaling argument for the solution 
$\phi_\rho(t,x):=\rho^{2/(p-1)}\phi(\rho t,\rho x)$ for $\rho\in \mathbb{R}$ 
yields the critical exponent of $p(s):=1+4/({3}-2s)$ 
for the well-posedness of the Cauchy problem of the equation 
in the Sobolev space $H^s(\mathbb{R}^{{3}})$ for $s\in \mathbb{R}$.
Especially, $p(-{3/2})$ is called the Fujita exponent, 
$p(0)$ is called the conformal exponent, 
$p(1)$ is called the energy critical exponent.
This exponent $p(s)$ also plays crucial  
roles for the Klein-Gordon equation, 
and it satisfies 
\[
1<p\left(-\frac{3}{2}\right)=\frac{5}{3}
<2<p(0)=\frac{7}{3}<p\left(\frac{1}{2}\right)=3<4<p(1)=5<6.
\]
By the simulations for $p=2,3,4,5,6$, 
we are able to know the behaviors of the solutions for $p$ which is equivalent or close to these critical exponents.

\subsection{Case 1: $H=0$}

To compare with the numerical results of SPS, CNS and RKS, we set the Hubble
constant $H$ as zero in this subsection.
In this case, $H_C$ is consistent with $E$, and $H_C$ is constant with time
evolutions theoretically.
To show the accuracy of the simulations, we monitor $H_C$.
\begin{figure}[htbp]
  \centering
  \includegraphics[keepaspectratio=true,width=\hsize]{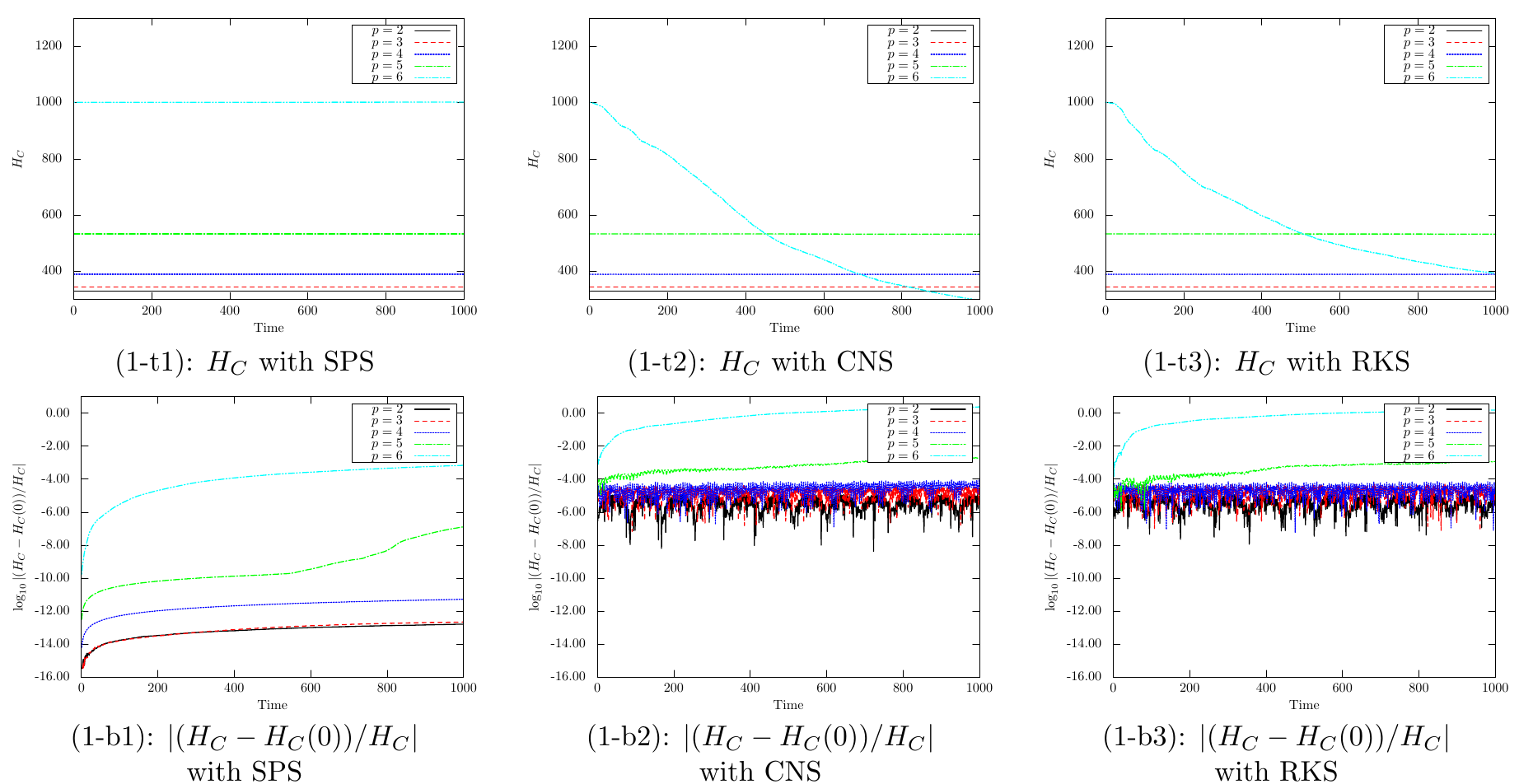}
  %{\small \textit{ 
  \caption{\label{fig:H0-A4p-H}
    The Hamiltonian $H_C$, and the difference between $H_C$ and
    the initial data $H_C(0)$ for $A=4$ and $\lambda=1$.
    The horizontal axis is time.
    The top panels are the values of $H_C$ and
    the bottom panels are the values of $\log_{10}|(H_C-H_C(0))/H_C|$.
    The left panels are drawn with SPS, the middle panels are with CNS,
    and the right panels are with RKS.
    %}}
  }
\end{figure}
We draw $H_C$, and the differences between $H_C$ and the initial value $H_C(0)$
in the evolutions with SPS, CNS and RKS for $A=4$ and $\lambda=1$ in
Fig.\ref{fig:H0-A4p-H}.
From the bottom panels (1-b1)--(1-b3), we see the values of
$|(H_C-H_C(0))/H_C|$ with CNS and RKS are larger than the ones with SPS.

\begin{figure}[htbp]
  \includegraphics[width=\hsize]{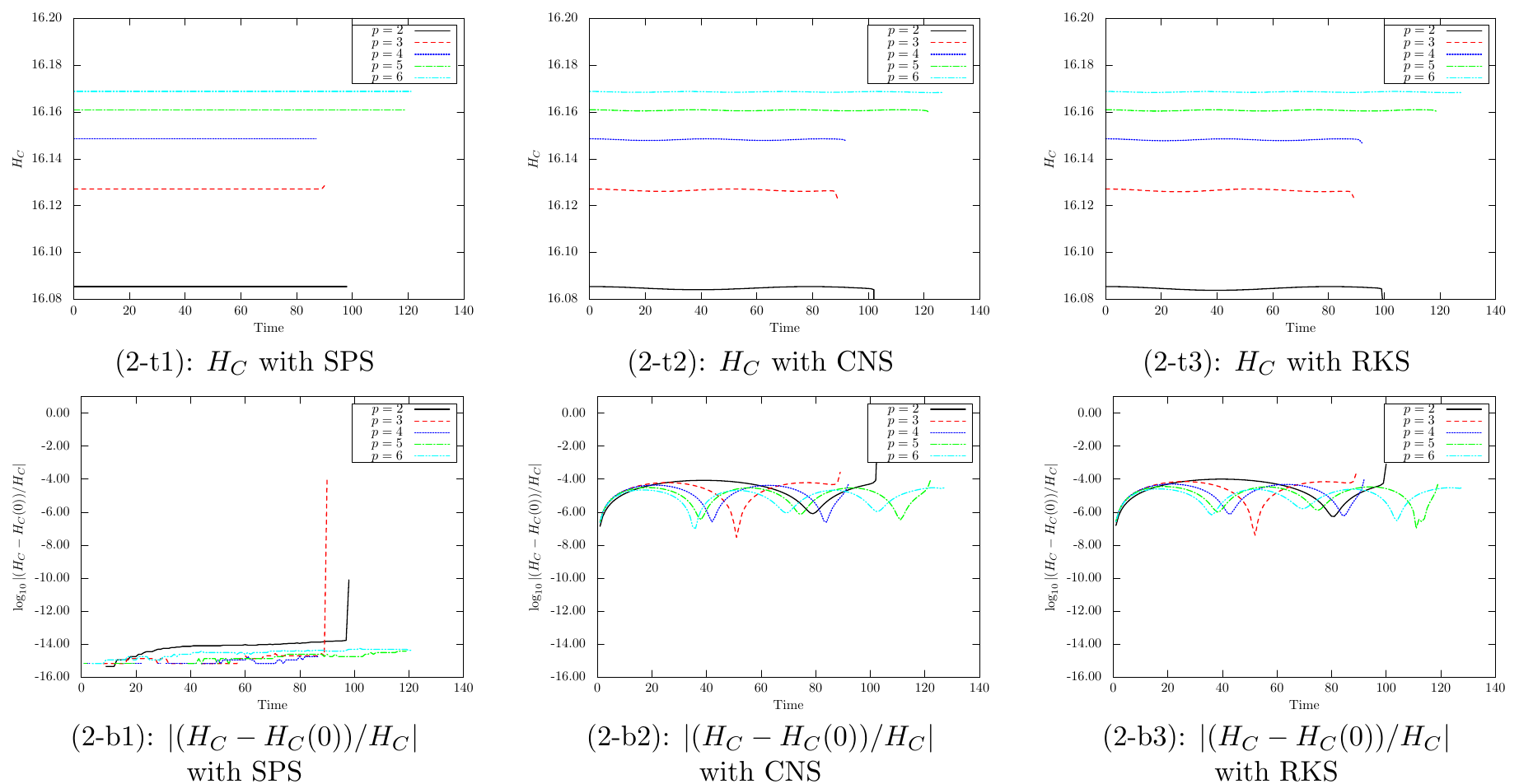}
%{\small \textit{ 
  \caption{\label{fig:H0-A09m-H}
    The same as Fig.\ref{fig:H0-A4p-H} except for the values of $A$ and
    $\lambda$.
    These results are set as $A=0.9$ and $\lambda=-1$.
    These calculations stop before $t=140$.
%}}
    }
\end{figure}

Next, we show the results for $A=0.9$ and $\lambda=-1$ in
Fig.\ref{fig:H0-A09m-H}.
We see the simulations stop before $t=140$.
From the bottom panels, we see the values of $|(H_C-H_C(0))/H_C|$ with SPS
are smaller than the ones with CNS and RKS.
In comparison with Fig.\ref{fig:H0-A09m-H}, the simulations in
Fig.\ref{fig:H0-A4p-H} are robust.
Since it is necessary condition of performing correct simulations that the
value of $|(H_C-H_C(0))/H_C|$ is small, KGE have to be performed with SPS.
Therefore, we use the numerical scheme as SPS hereafter.
\begin{figure}[htbp]
  \includegraphics[width=\hsize]{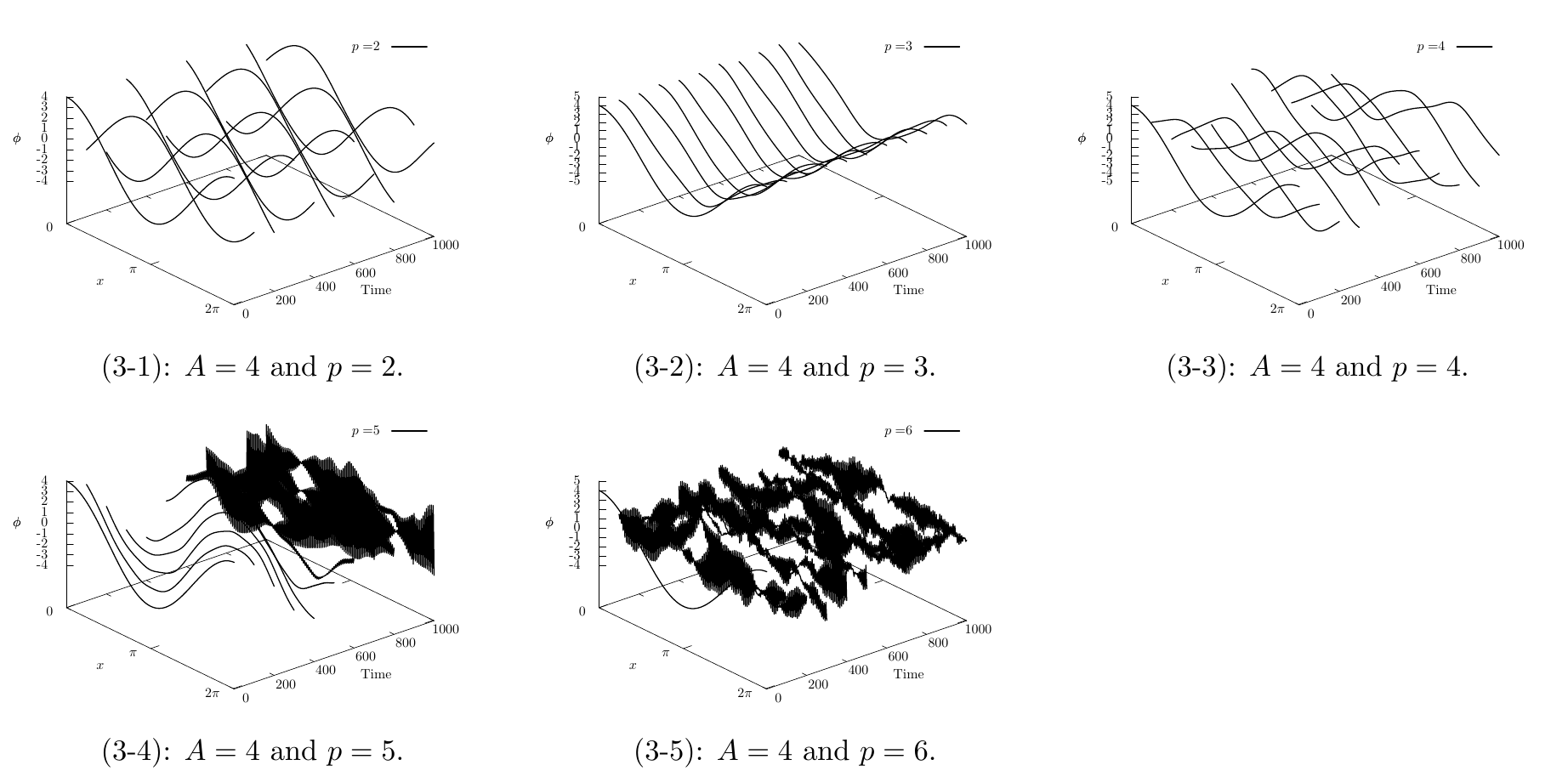}
%{\small \textit{ 
  \caption{\label{fig:H0-A4p-phi}
    $\phi$ for $A=4$ and $\lambda=1$ with SPS.
    The vibrations for $p=5, 6$ which are Panel (3-4) and Panel (3-5) occur
    after $t=500$ and $t=100$, respectively.
%}}
  }
\end{figure}
\begin{figure}[htbp]
  \includegraphics[width=\hsize]{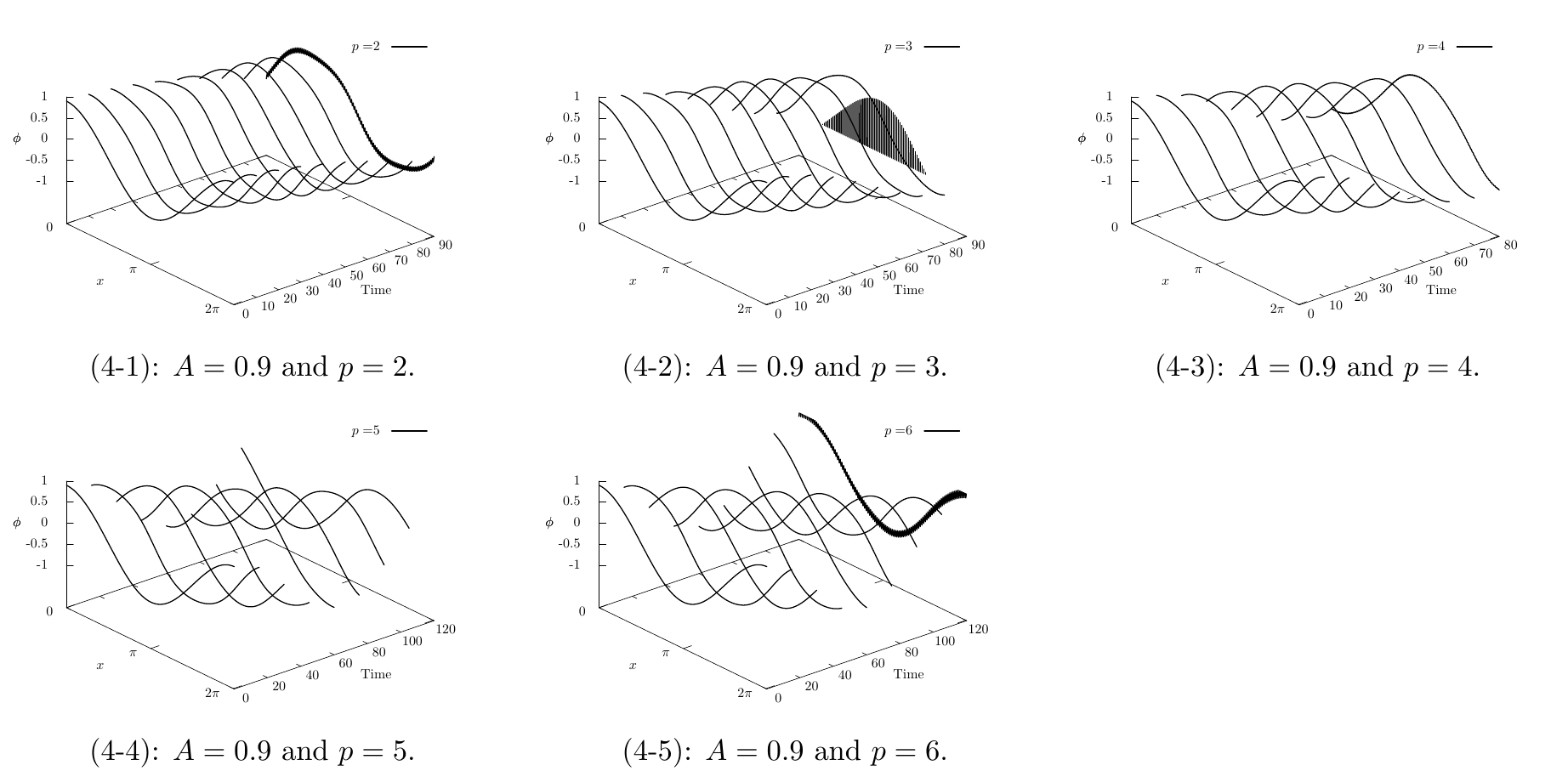}
%{\small \textit{ 
  \caption{\label{fig:H0-A09m-phi}
    The same as Fig.\ref{fig:H0-A4p-phi} except for the values of $A$ and
    $\lambda$.
    These results are set as $A=0.9$ and $\lambda=-1$.
    All of the simulations stop before $t=140$.
%}}
  }
\end{figure}
The results of $\phi$ with SPS are shown in Fig.\ref{fig:H0-A4p-phi} and
Fig.\ref{fig:H0-A09m-phi}.
Fig.\ref{fig:H0-A4p-phi} shows the case of $A=4$ and $\lambda=1$.
We see that the vibrations occur for $p=5, 6$.
Fig.\ref{fig:H0-A09m-phi} shows the case of $A=0.9$ and $\lambda=-1$.
We see all of the simulations stop before $t=140$.

\subsection{Case 2: $H=10^{-3}$}

In this subsection and next subsection, we \textcolor{red}{perform} the
numerical \textcolor{red}{simulations} for $H>0$.
First, we calculate for $\lambda=1$ and $H=10^{-3}$.
When $H\neq0$, $H_C$ is not constant in time evolutions.
Since we cannot judge the reliability of simulations by monitoring the values
of $H_C$, we show $\tilde{H}_C$ defined by \eqref{eq:modTotalH} instead of
$H_C$.
\begin{figure}[htbp]
  \centering
  \includegraphics[width=\hsize]{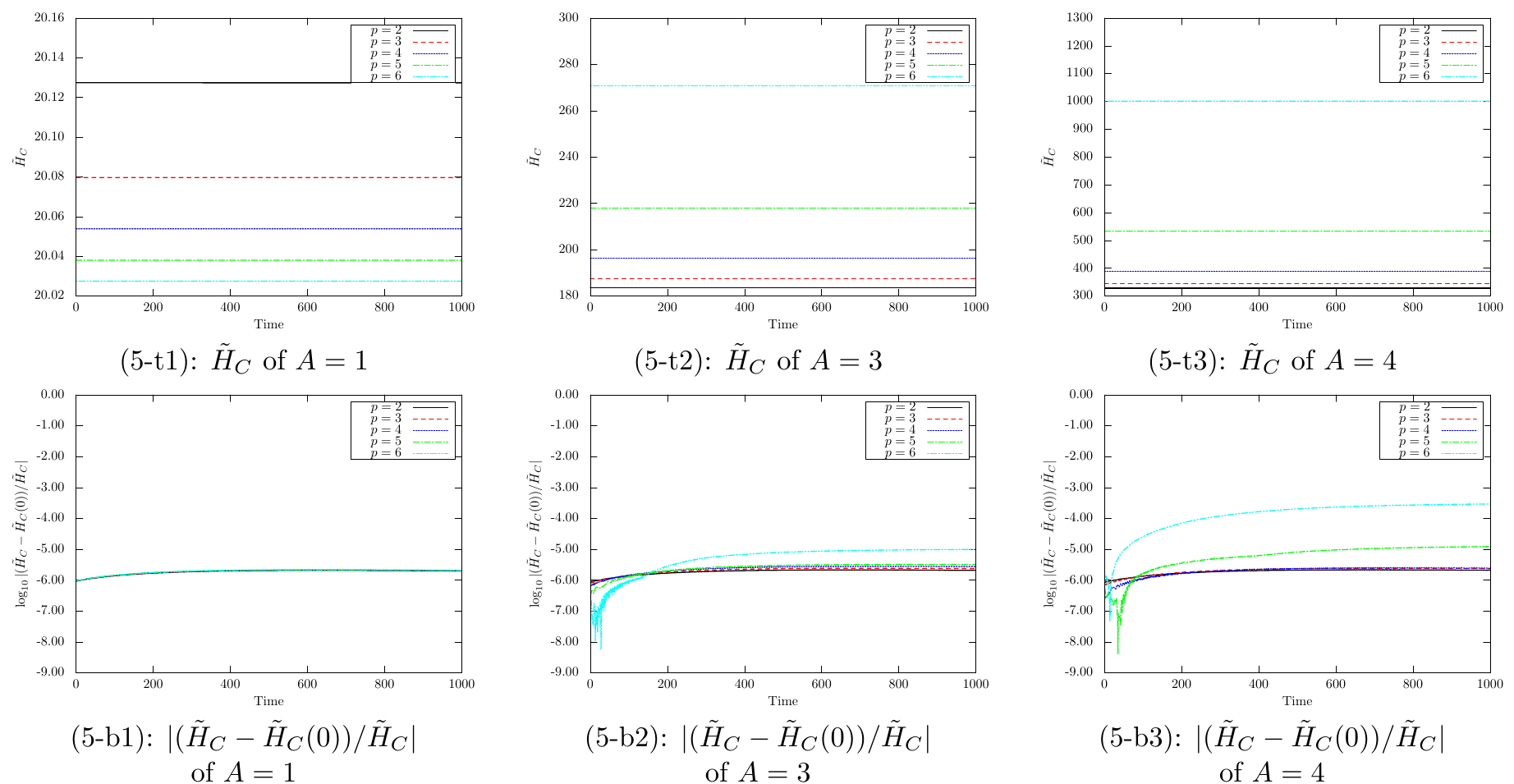}
%{\small \textit{ 
  \caption{\label{fig:He-3p-H}
    $\tilde{H}_C$, and the difference between $\tilde{H}_C$ and the
    initial data $\tilde{H}_C(0)$ for $A=1,3,4$, $\lambda=1$ and $H=10^{-3}$.
    The horizontal axis is time.
    The top panels show
    the values of $\tilde{H}_C$, and
    the bottom panels show the values of
    $\log_{10}|(\tilde{H}_C-\tilde{H}_C(0))/\tilde{H}_C|$.
    The left panels are the case of $A=1$, the middle panels are $A=3$,
    and the right panels are $A=4$.
    The lines of Panel (5-b1) are almost overlapping.
%}}
  }
\end{figure}
Fig.\ref{fig:He-3p-H} shows $\tilde{H}_C$, and the differences between
$\tilde{H}_C$ and the initial value $\tilde{H}_C(0)$ for $A=1, 3, 4$.
We see the value of $p=2$ is 
the largest and that of $p=6$ is 
the smallest
in the panel (5-t1).
Contrarily, in the others of the top panels we see the largest value is the
case of $p=6$.
The reason is $A=1$ or not.
If $A=1$, the nonlinear term becomes small as $p$ becomes large.
On the other hand, if $A>1$, the value becomes large as $p$ becomes large.
From the bottom panels (5-b1)--(5-b3), we see
$|(\tilde{H}_C-\tilde{H}_C(0))/\tilde{H}_C|$ of the case $p=6$ is
the largest of them in each panel but all of the values in the bottom panels are
enough small.
\begin{figure}[htbp]
  \includegraphics[width=\hsize]{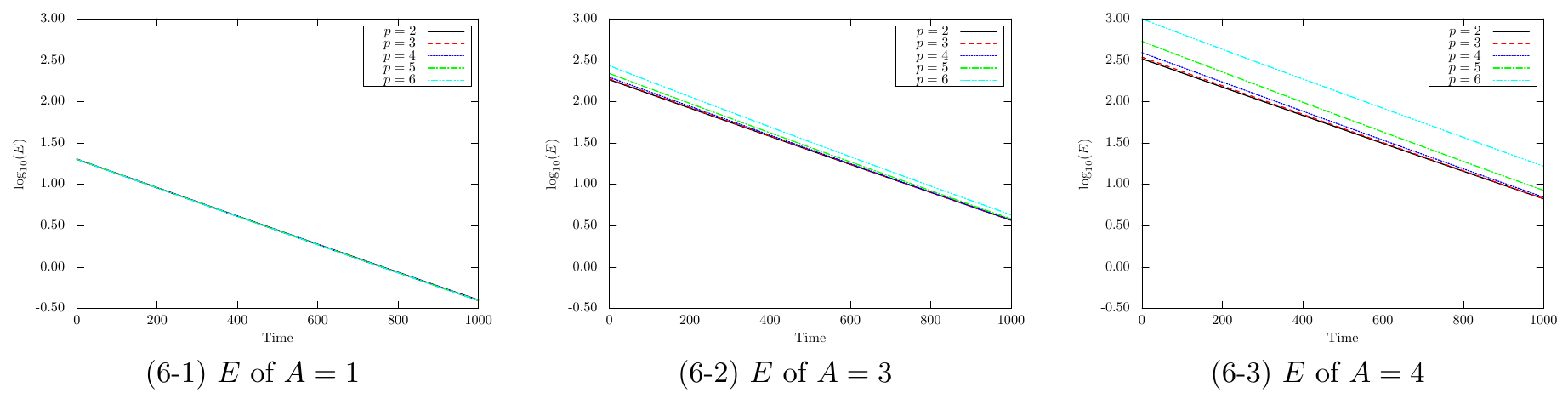}
%{\small \textit{ 
  \caption{\label{fig:He-3p-E}
    The total energy $E$ for $\lambda=1$ and $H=10^{-3}$.
    The horizontal axis is time, and the vertical axis is $\log_{10}(E)$.
    The left panel is the case of $A=1$,
    the middle panel is $A=3$ and
    the right panel is $A=4$.
    The lines of Panel (6-1) are almost overlapping.
%}}
  }
\end{figure}
Fig.\ref{fig:He-3p-E} shows the total energy $E$ define{d} by \eqref{eq:totalE}.
We see that the energy is exponential decay.
Furthermore, the slopes of the lines in the panels of the figure are
almost the same.
This result means the diffusion effect with time evolutions is not caused
by changing in $A$ or $p$.
While the lines in Panel (6-1) which is the case of $A=1$ are almost
overlapping, there are differences between the values of $E$ for $A=3, 4$.
To investigate the reason, we show the components of $E$ defined by
\eqref{eq:componentsE}.
\begin{figure}[htbp]
  \centering
  \includegraphics[width=\hsize]{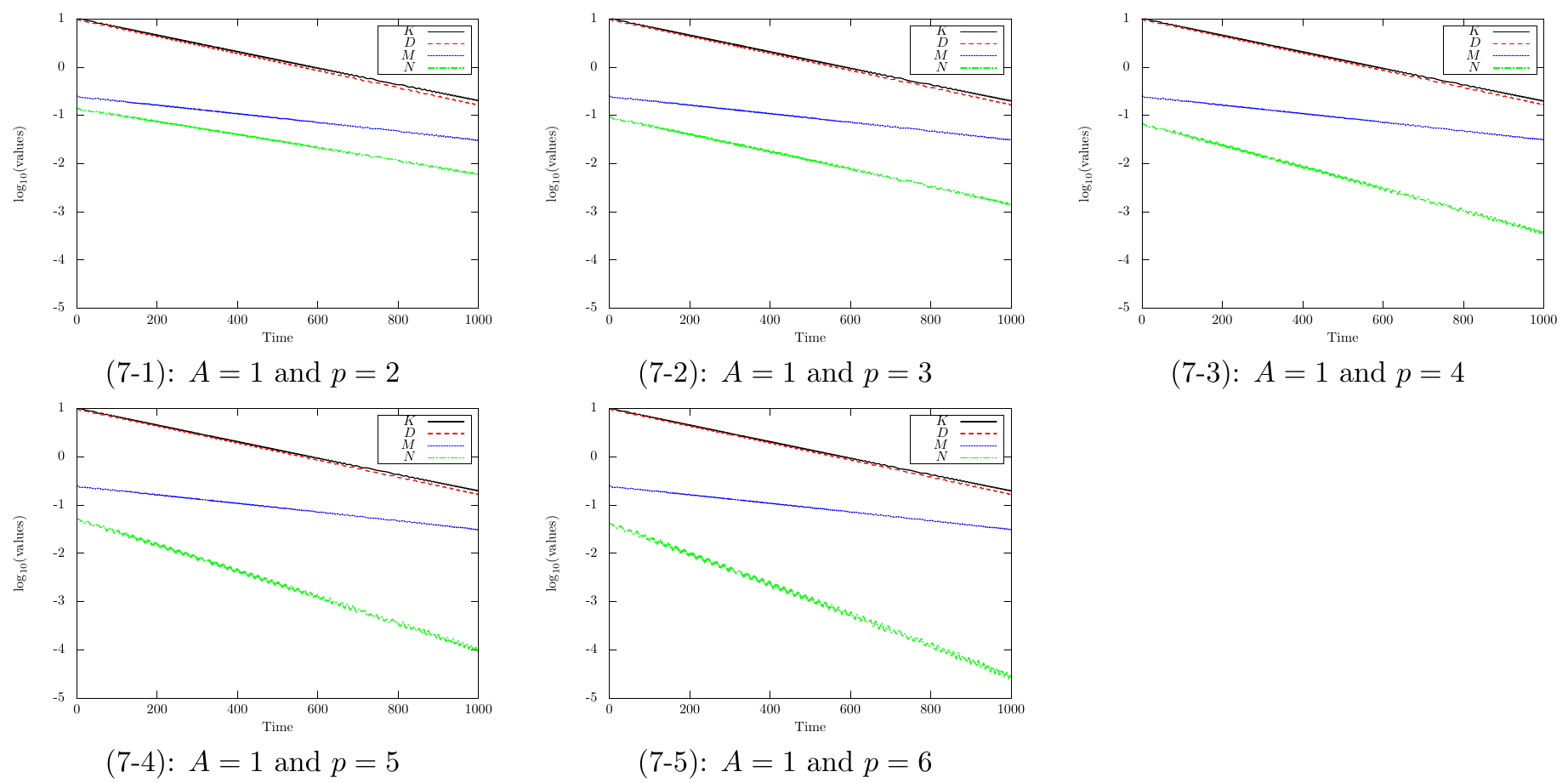}
%{\small \textit{ 
  \caption{\label{fig:He-3-A1p-terms}
    The kinetic term $K$, the diffusion term $D$, the mass term $M$, and the
    nonlinear term $N$ are shown for $A=1$, $\lambda=1$ and $H=10^{-3}$.
    The horizontal axis is time and the vertical axis is the logarithm{ic} values
    of $K$, $D$, $M$ and $N$.
%}}
  }
\end{figure}
\begin{figure}[htbp]
  \centering
  \includegraphics[width=\hsize]{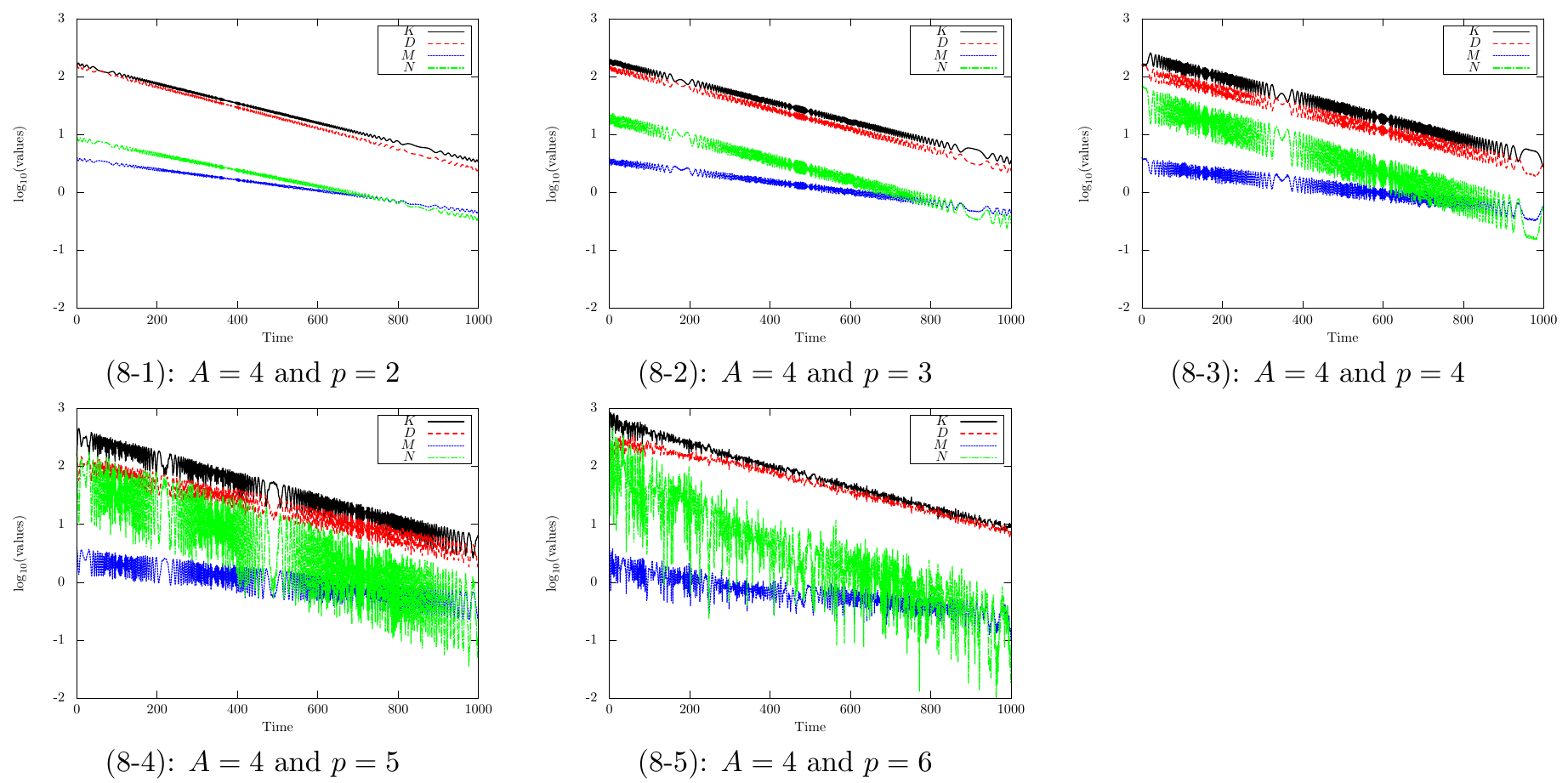}
%{\small \textit{ 
  \caption{\label{fig:He-3-A4p-terms}
    The same as Fig.\ref{fig:He-3-A1p-terms} except for the value of $A$.
    These results are set as $A=4$.
%}}
  }
\end{figure}

The kinetic term $K$, the diffusion term $D$, the mass term $M$ and the
nonlinear term $N$ for $A=1$ are shown in Fig.\ref{fig:He-3-A1p-terms}
and the ones for $A=4$ are in Fig.\ref{fig:He-3-A4p-terms}.
We see that $K$ and $D$ are the dominant terms, and $M$ and $N$ are very small
compared with $K$ and $D$ in Fig.\ref{fig:He-3-A1p-terms}.
  The decline of the energy-component $N$ is becoming faster than $K$, $D$ and
  $M$ 
  as the power $p$ increases.
On the other hand, the proportion of $N$ to $E$ becomes large as $p$
becomes large in
Fig.\ref{fig:He-3-A4p-terms}.
This is because there are the differences between $E$s 
in Fig.\ref{fig:He-3p-E}.
\begin{figure}[htbp]
  \centering
  \includegraphics[width=\hsize]{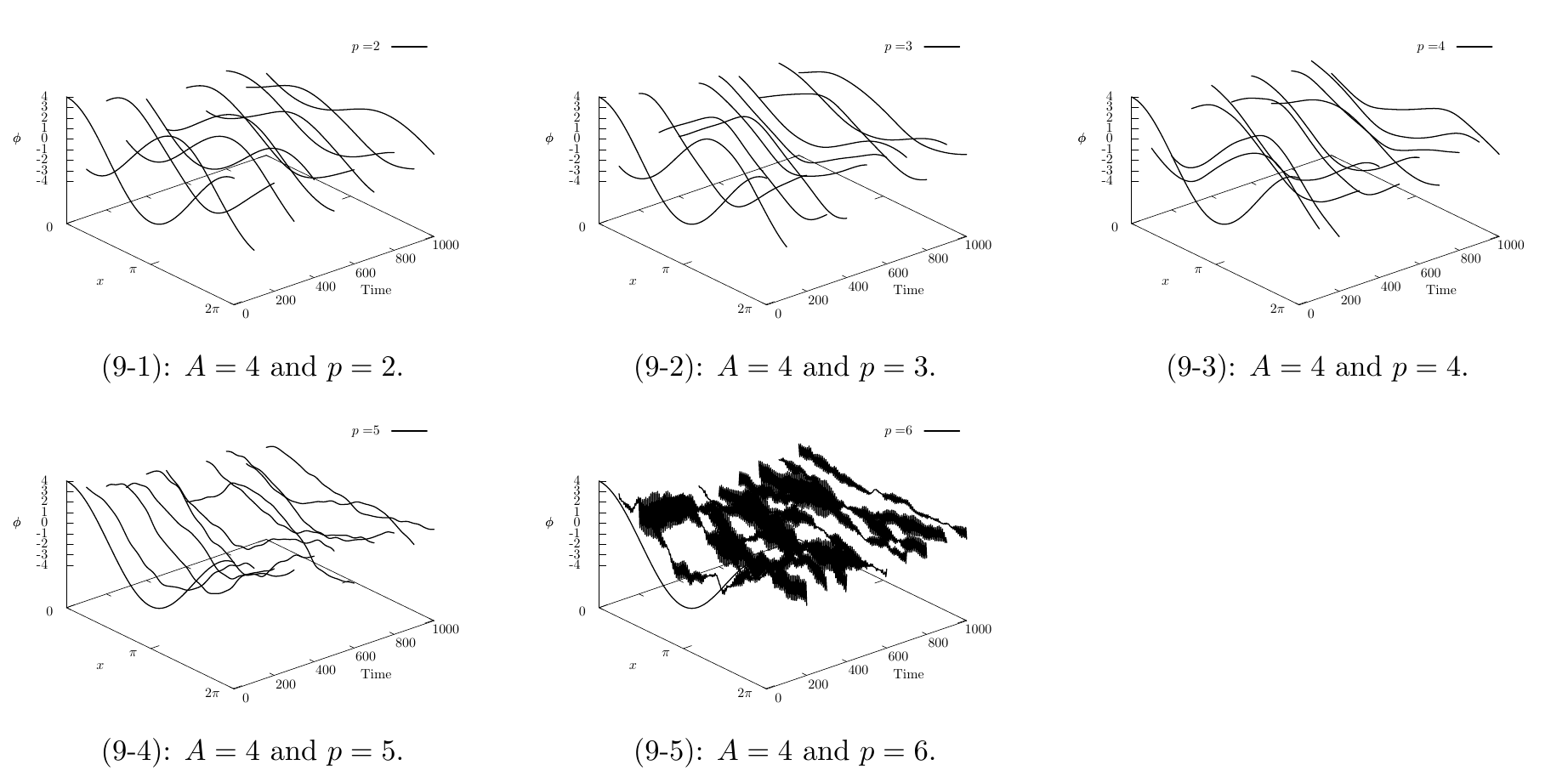}
%{\small \textit{ 
  \caption{\label{fig:He-3-A4p-phi}
    The same as Fig.\ref{fig:H0-A4p-phi} except for the value of $H$.
    These results are set as $H=10^{-3}$.
    The vibrations occur for $p=6$ which is Panel (9-5).
%}}
  }
\end{figure}
Fig.\ref{fig:He-3-A4p-phi} shows $\phi$ for $A=4$ and $\lambda=1$, and the
numerical settings are the same as Fig.\ref{fig:H0-A4p-phi} except
for the value of $H$.
The vibrations occur in the only case of $p=6$.
Since the vibrations occur in the cases of $p=5,6$ in
Fig.\ref{fig:H0-A4p-phi}, the Hubble constant seems to affect generations of
the vibrations of $\phi$.

Next we show the results of the simulations with the case of $\lambda=-1$.
\begin{figure}[htbp]
  \centering
  \includegraphics[width=\hsize]{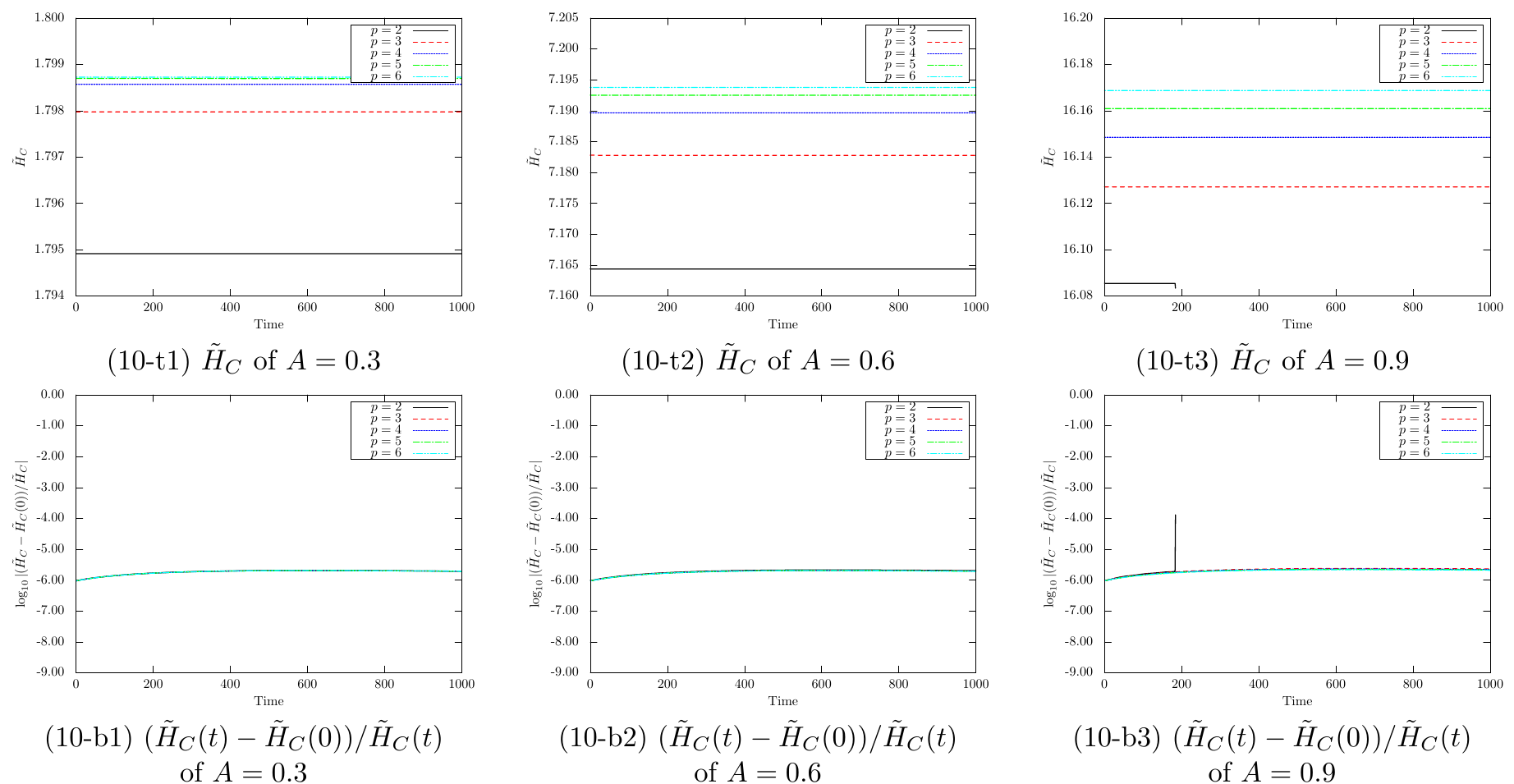}
%{\small \textit{ 
  \caption{\label{fig:He-3m-H}
    The same as Fig.\ref{fig:He-3p-H} except for the values of $A$ and
    $\lambda$.
    These results are set as $A=0.3, 0.6, 0.9$ and $\lambda=-1$.
    The simulation for $p=2$ and $A=0.9$ stops before $t=200$.
%}}
  }
\end{figure}
Fig.\ref{fig:He-3m-H} shows $\tilde{H}_C$ and $|(\tilde{H}_C-\tilde{H}_C(0))/
\tilde{H}_C|$ for $A=0.3, 0.6, 0.9$, $\lambda=-1$ and $H=10^{-3}$.
While we see the simulation for $A=0.9$ and $p=2$ stops before $t=200$ 
{in} 
Panel (10-t3) and Panel (10-b3),
{the values }
$|(\tilde{H}_C-\tilde{H}_C(0))/\tilde{H}_C|$ for the other cases are enough
small.
\begin{figure}[htbp]
  \includegraphics[width=\hsize]{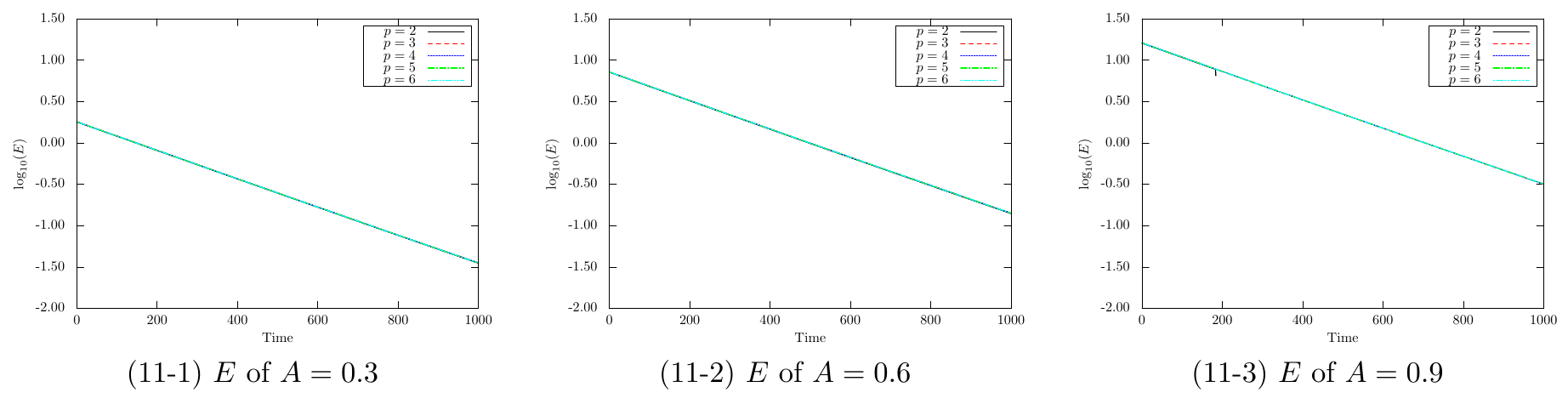}
%{\small \textit{ 
  \caption{\label{fig:He-3m-E}
    The same as Fig.\ref{fig:He-3p-E} except for the values of $A$ and
    $\lambda$.
    These results are calculated for $A=0.3, 0.6, 0.9$ and $\lambda=-1$.
    The lines are almost overlapping in each panel except for the case of $p=2$
    and $A=0.9$.
    The simulation of $p=2$ and $A=0.9$ stops before $t=200$.
%}}
  }
\end{figure}
Fig.\ref{fig:He-3m-E} shows $E$ for $A=0.3, 0.6, 0.9$, $\lambda=-1$ and
$H=10^{-3}$.
It seems there is no difference between the slopes of all the lines in
Fig.\ref{fig:He-3p-E} and Fig.\ref{fig:He-3m-E} 
{regardless of the values of $A$, $p$ and $\lambda$}.
Thus, the dissipation effect of $E$ is not caused by changing in $A$, $p$ and
the signature of $\lambda$.
\begin{figure}[htbp]
  \centering
  \includegraphics[width=\hsize]{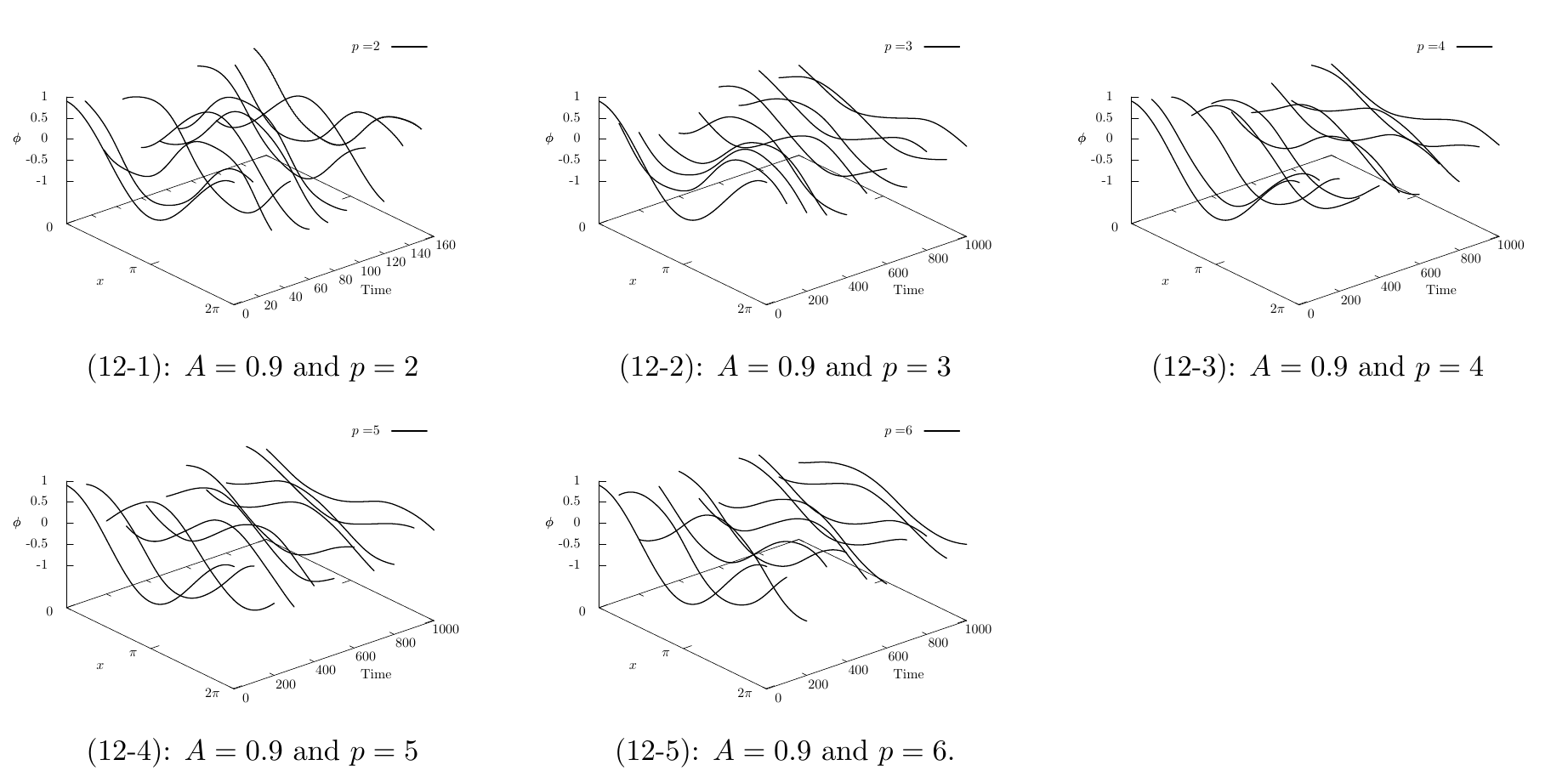}
%{\small \textit{ 
  \caption{\label{fig:He-3-A09m-phi}
    The same as Fig.\ref{fig:H0-A09m-phi} except for the values of $H$.
    These results are set as $H=10^{-3}$.
    The simulation of the case $p=2$ stops before $t=180$.
%}}
  }
\end{figure}
Fig.\ref{fig:He-3-A09m-phi} shows $\phi$ for $A=0.9$, $\lambda=-1$ and
$H=10^{-3}$.
These results are the same as Fig.\ref{fig:H0-A09m-phi} except for the value of
$H$.
By comparing Fig.\ref{fig:H0-A09m-phi} 
with Fig.\ref{fig:He-3-A09m-phi}, the
simulation times for $H=10^{-3}$ are longer than that for $H=0$.
It means the Hubble constant would influence the stability of the
simulation.
By comparing Fig.\ref{fig:He-3-A4p-phi} and Fig.\ref{fig:He-3-A09m-phi}, the
simulations for $\lambda=1$ are robust against the case of $\lambda=-1$
since the calculation time for $p=2$ and $\lambda=-1$ is short.
The tendency is the same as the case of $H=0$.

\subsection{Case 3: $H=10^{-2}$}

In this subsection, we perform some simulations with $H=10^{-2}$ and
$\lambda=1$.
\begin{figure}[htbp]
  \centering
  \includegraphics[width=\hsize]{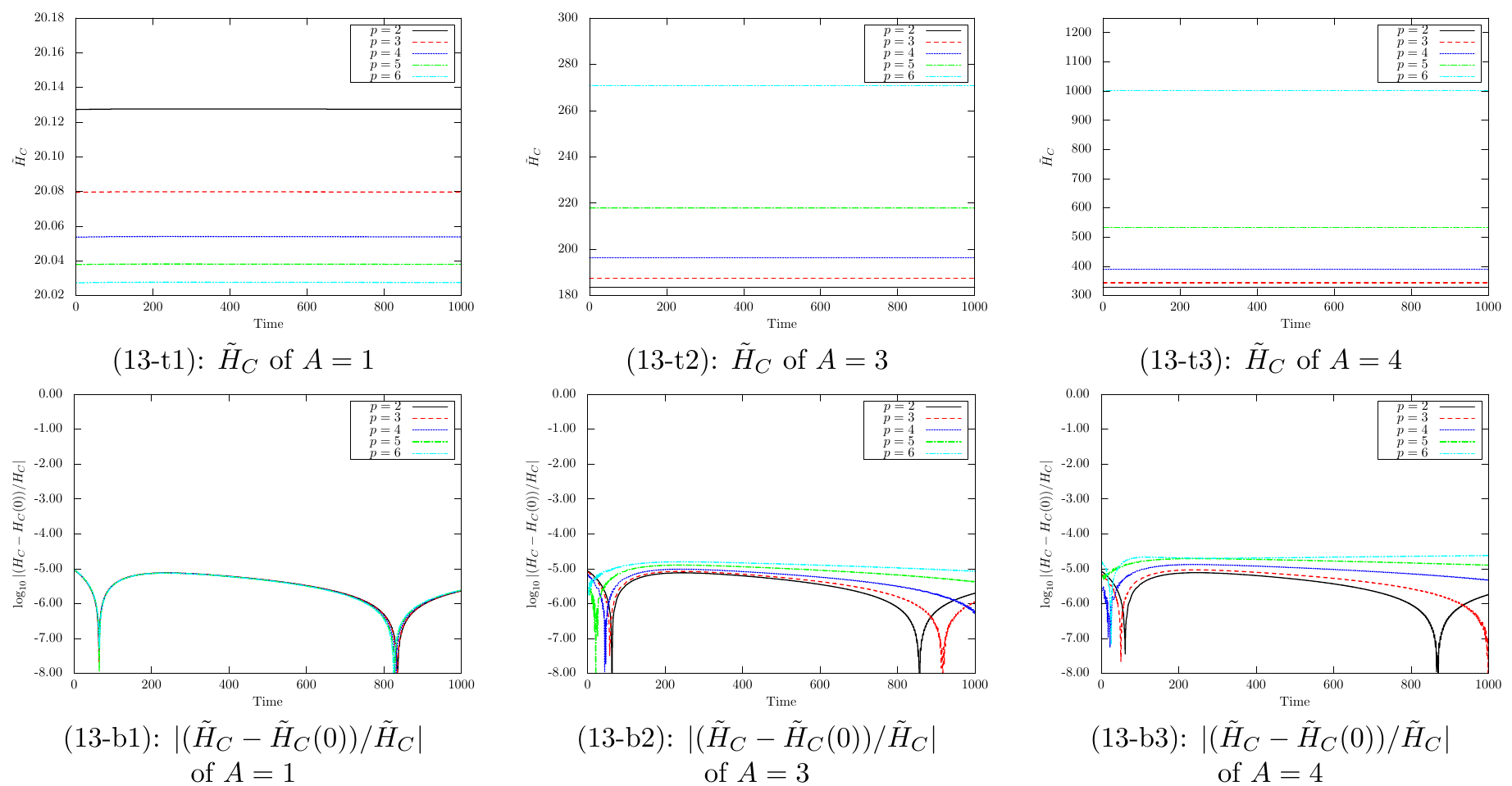}
%{\small \textit{ 
  \caption{\label{fig:He-2p-H}
    The same as Fig.\ref{fig:He-3p-H} except of the Hubble constant $H$.
    These results are set as $H=10^{-2}$.
    The lines in Panel (13-b1) are almost overlapping.
%}}
  }
\end{figure}
Fig.\ref{fig:He-2p-H} shows $\tilde{H}_C$ and $|(\tilde{H}_C-\tilde{H}_C(0))
/\tilde{H}_C|$ for $H=10^{-2}$ and this figure is the same as
Fig.\ref{fig:He-3p-H} except for the value of $H$.
We see all the values of $|(\tilde{H}_C-\tilde{H}_C(0))/\tilde{H}_C|$ for
$H=10^{-2}$ are enough small.
\begin{figure}[htbp]
  \includegraphics[width=\hsize]{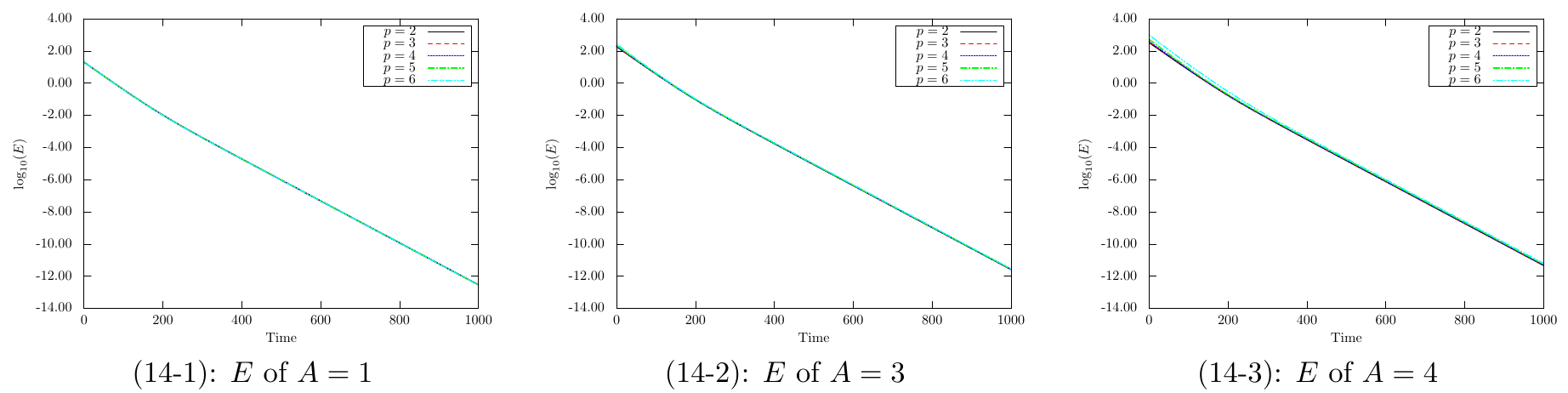}
%{\small \textit{ 
  \caption{\label{fig:He-2p-E}
    The same as Fig.\ref{fig:He-3p-E} except for the Hubble constant $H$.
    These results are set as $H=10^{-2}$.
    The lines in Panel (14-1) and Panel (14-2) are almost overlapping in each
    panel.
%}}
  }
\end{figure}
Fig.\ref{fig:He-2p-E} shows $E$ with the Hubble constant as $10^{-2}$.
There are few differences between the changes of $p$ 
compared with the cases of $H=10^{-3}$ in Fig.\ref{fig:He-3p-E}.
By comparing Fig.\ref{fig:He-3p-E} and Fig.\ref{fig:He-2p-E}, the diffusion
effect of $E$ for $H=10^{-2}$ is stronger than that for $H=10^{-3}$.
Thus, the diffusion effect is mainly caused by the Hubble constant, and
the effect would become strong as $H$ becomes large.
\begin{figure}[htbp]
  \includegraphics[width=\hsize]{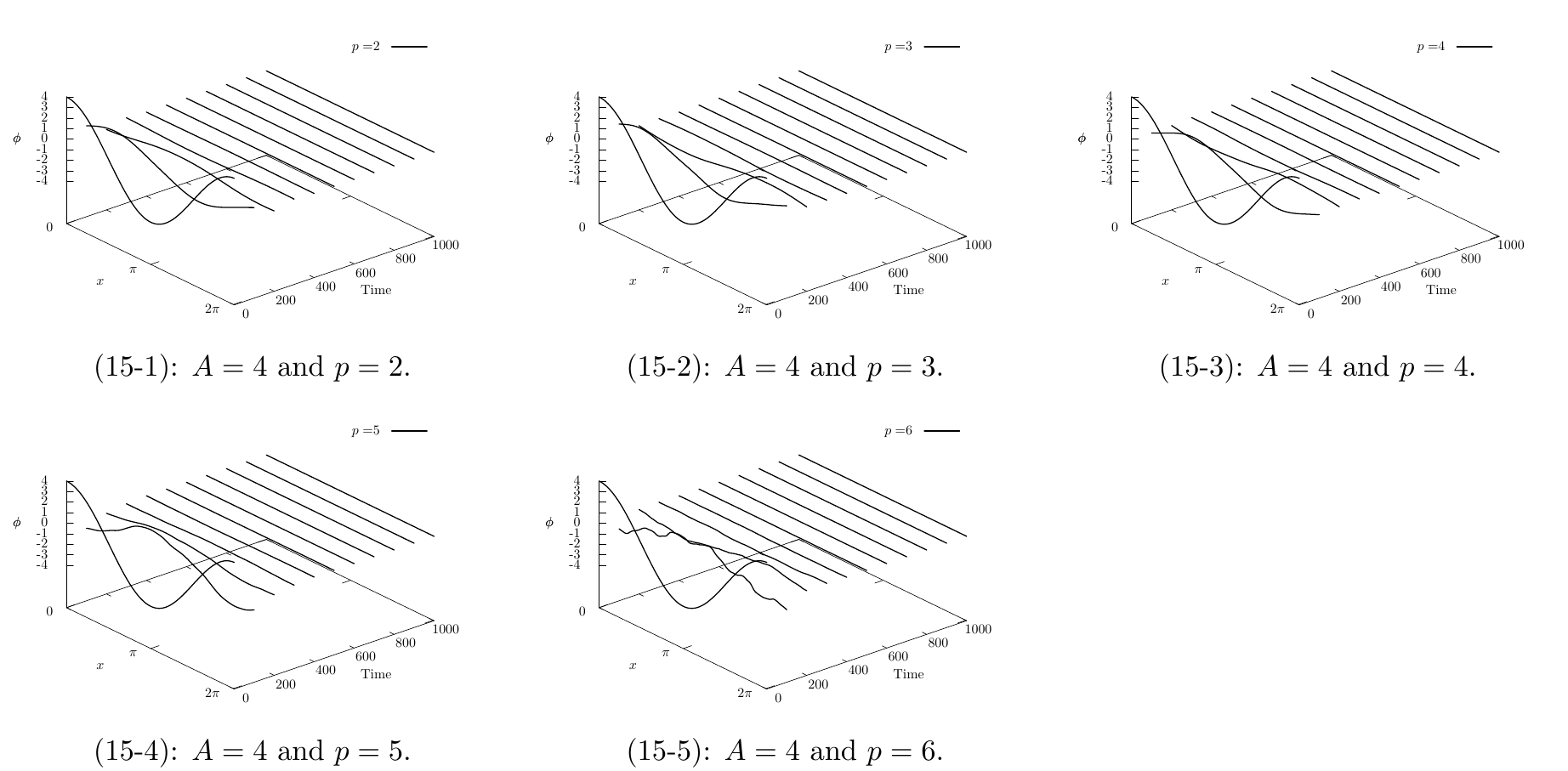}
%{\small \textit{ 
  \caption{\label{fig:He-2-A4p-phi}
    The same as Fig.\ref{fig:H0-A4p-phi} and Fig.\ref{fig:He-3-A4p-phi} except
    for the Hubble constant $H$.
    These results are set as $H=10^{-2}$.
%}}
  }
\end{figure}
In Fig.\ref{fig:He-2-A4p-phi} which shows $\phi$ for $H=10^{-2}$, we see the
waveform is damped and there are few vibrations of the waveform.
By comparing Fig.\ref{fig:H0-A4p-phi}, Fig.\ref{fig:He-3-A4p-phi} and
Fig.\ref{fig:He-2-A4p-phi},
the stability of the simulations becomes good as
$H$ become{s} large since the vibrations of $\phi$ 
decreases as $H$ becomes large.

Next we show the results for $\lambda=-1$.
\begin{figure}[htbp]
  \centering
  \includegraphics[width=\hsize]{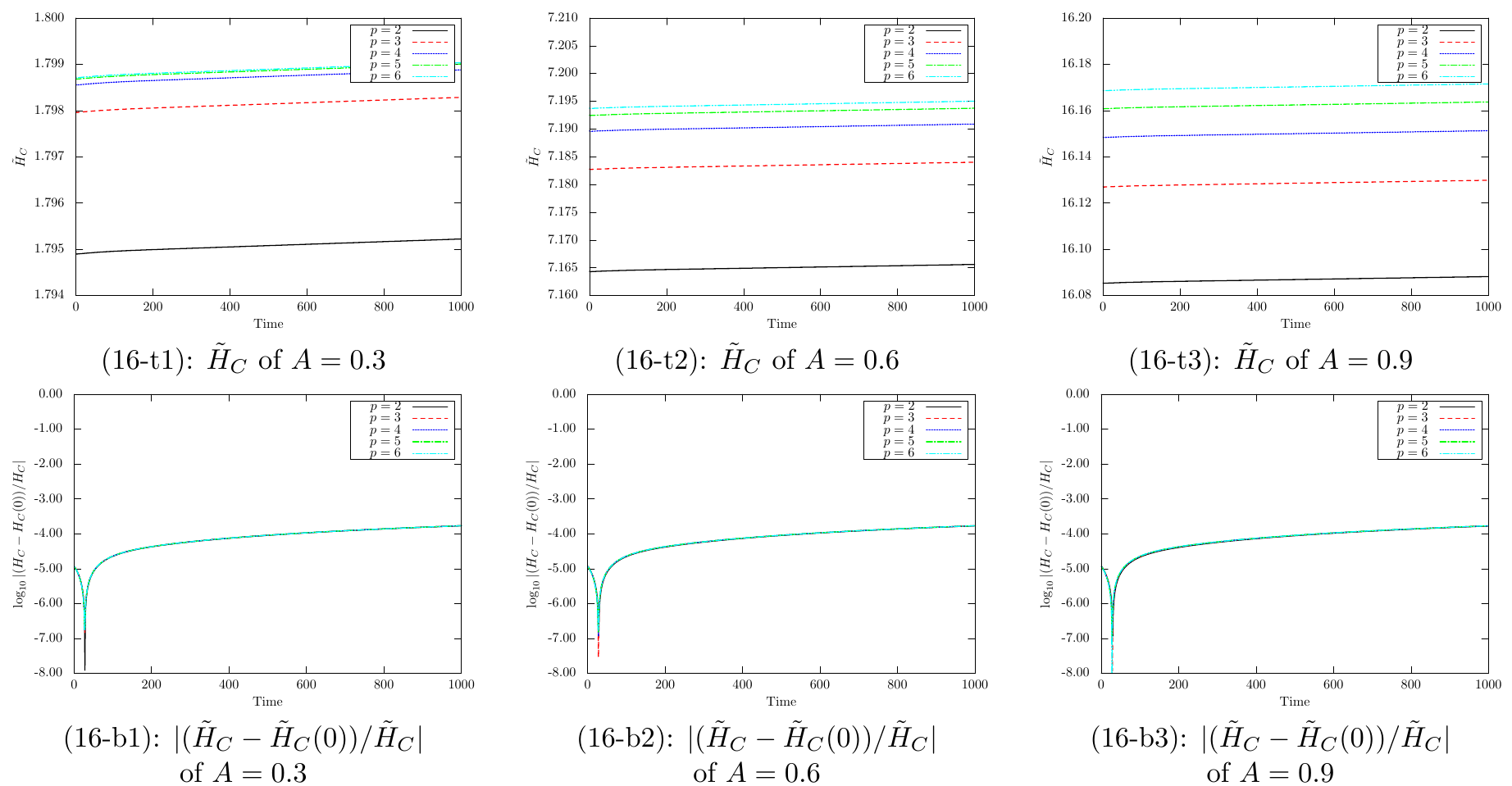}
%{\small \textit{ 
  \caption{\label{fig:He-2m-H}
    The same as Fig.\ref{fig:He-2p-H} except {for} the value of $\lambda$.
    These results are set as $\lambda=-1$.
    The lines in the bottom panels are almost overlapping each other.
%}}
  }
\end{figure}
In Fig.\ref{fig:He-2m-H}, we see the simulations are reliable 
since all of the
values of $|(\tilde{H}_C-\tilde{H}_C(0))/\tilde{H}_C|$ are enough small.
\begin{figure}[htbp]
  \includegraphics[width=\hsize]{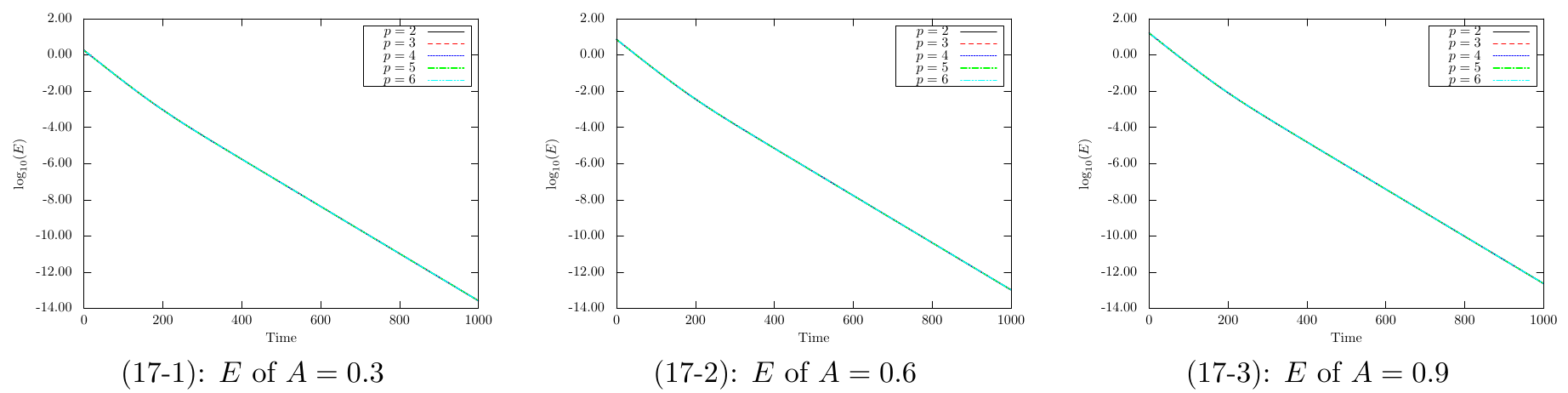}
%{\small \textit{ 
  \caption{\label{fig:He-2m-E}
    The same as Fig.\ref{fig:He-3m-E} except for $H$.
    These results are set as $H=10^{-2}$.
    The lines are almost overlapping each other.
%}}
  }
\end{figure}
Fig.\ref{fig:He-2m-E} shows the total energy $E$.
By comparing Fig.\ref{fig:He-3m-E} and Fig.\ref{fig:He-2m-E},
we see the diffusion effect for $\lambda=-1$ and $H=10^{-2}$ is stronger than
that for $\lambda=-1$ and $H=10^{-3}$.
Thus, considering the results for $\lambda=1$, the diffusion effect is mainly
caused by not the signature of the nonlinear term but the Hubble constant.
\begin{figure}[htbp]
  \centering
  \includegraphics[width=\hsize]{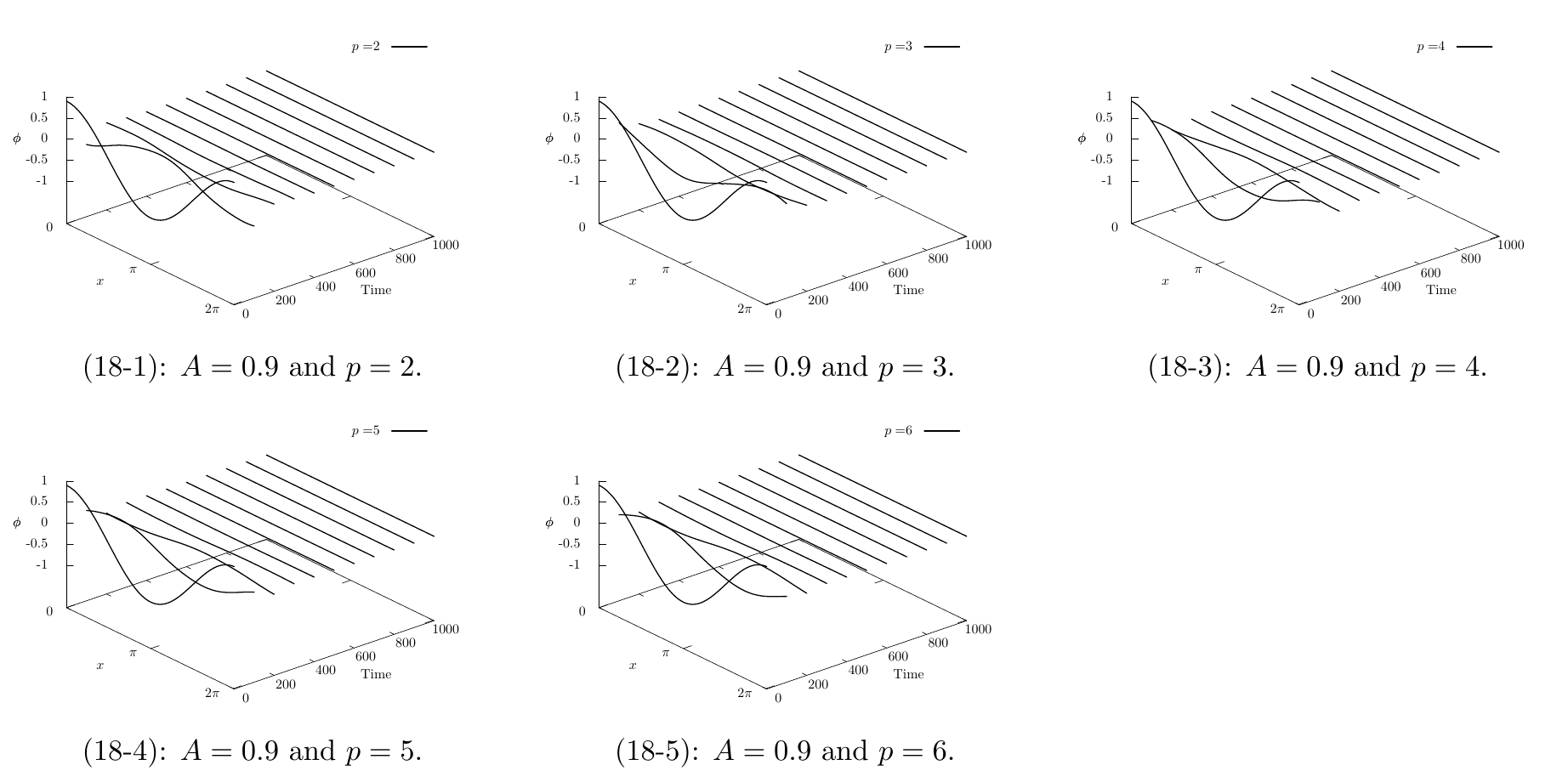}
%{\small \textit{ 
  \caption{\label{fig:He-2-A09m-phi}
    The same as Fig.\ref{fig:H0-A09m-phi} and Fig.\ref{fig:He-3-A09m-phi}
    except for $H$.
    These results are set as $H=10^{-2}$.
%}}
  }
\end{figure}
Fig.\ref{fig:He-2-A09m-phi} shows $\phi$ for $\lambda=-1$ and $H=10^{-2}$.
The simulation times become long as $H$ becomes large from the results in
Fig.\ref{fig:H0-A09m-phi}, Fig.\ref{fig:He-3-A09m-phi} and
Fig.\ref{fig:He-2-A09m-phi}.
These results indicate the simulations become stable as $H$ becomes large.

\newsection{Concluding remark}
In this paper, we made the Hamiltonian formulation of the semilinear
Klein-Gordon equation{,} and {we derived } the discrete equation with the structure-preserving
scheme (SPS).
To show the reliability of the simulations, we proposed {the} constant value
$H_C$ when the Hubble constant is zero case and $\tilde{H}_C$ when the Hubble
constant is nonzero case.
With SPS, the Crank-Nicolson scheme (CNS) and the Runge-Kutta scheme (RKS), 
we performed some simulations in flat spacetime.
Then, we showed the superiority of SPS to CNS and RKS in some simulations.
We performed some simulations with small $\tilde{H}_C$ and
showed the influence of the Hubble constant on the numerical stability.
Especially, if the signature of the nonlinear term is negative, the simulations
stop in some cases.
However, with the negative nonlinear term, we showed the enough large value of
the Hubble constant gives the long and stable simulations.
It is remarkable that
the diffusion effect caused by the positive Hubble constant is much stronger
than
the nonlinear term.
Thus, we conclude that we are able to perform stable simulations when the
Hubble constant is a sufficiently large.
While the diffusion effects are expected from the theoretical point of view 
since the equation \eqref{KG-deSitter} has the positive-dissipative term $nH\partial_t\phi$ 
for the positive Hubble constant ($H>0$), 
the case of the negative Hubble constant ($H<0$) seems to be unstable 
and requires more delicate consideration 
since the numerical errors must be rigorously estimated  for the blow-up solutions 
in the unstable case,
which will be reported in the subsequent paper.

\section*{Acknowledgements}
This work was supported by JSPS KAKENHI Grant Number 16H03940 (M.N.).

%%%%%%%%%%%%%%%%%%%%%%%%%%%%%%%%%%
%
%%%%%%%%%%%%%%%%%%%%%%%%%%%%%%%%%%

\end{document}